\documentclass[twocolumn]{aastex62} 

\usepackage[ampersand]{easylist}
\usepackage{xspace}
\usepackage{color}
\usepackage{rotating}
\usepackage{longtable}
\usepackage{amsmath, cases}
\usepackage{subfigure}
\usepackage[ampersand]{easylist}
\PassOptionsToPackage{hyphens}{url}\usepackage{hyperref}
\usepackage{natbib}

\newcommand{\samurai}{\texttt{samurai}}
\newcommand{\emcee}{\texttt{emcee}}

\shorttitle{Glint Mapping}
\shortauthors{Lustig-Yaeger et al.}

\begin{document}

\title{Detecting Ocean Glint on Exoplanets Using Multiphase Mapping}

\correspondingauthor{Jacob Lustig-Yaeger}
\email{jlustigy@uw.edu}

\author{Jacob Lustig-Yaeger} 
\affiliation{Astronomy Department, University of Washington, Box 951580, Seattle, WA 98195, USA}
\affiliation{NASA Astrobiology Institute -- Virtual Planetary Laboratory Team, USA}
\affiliation{Astrobiology Program, University of Washington, 3910 15th Ave. NE, Box 351580, Seattle, WA 98195, USA}

\author{Victoria S. Meadows}
\affiliation{Astronomy Department, University of Washington, Box 951580, Seattle, WA 98195, USA}
\affiliation{NASA Astrobiology Institute -- Virtual Planetary Laboratory Team, USA}
\affiliation{Astrobiology Program, University of Washington, 3910 15th Ave. NE, Box 351580, Seattle, WA 98195, USA}

\author{Guadalupe Tovar Mendoza}
\affiliation{Astronomy Department, University of Washington, Box 951580, Seattle, WA 98195, USA}
\affiliation{NASA Astrobiology Institute -- Virtual Planetary Laboratory Team, USA}
\affiliation{Astrobiology Program, University of Washington, 3910 15th Ave. NE, Box 351580, Seattle, WA 98195, USA}

\author{Edward W. Schwieterman}
\affiliation{NASA Astrobiology Institute -- Virtual Planetary Laboratory Team, USA}
\affiliation{NASA Postdoctoral Fellow, Department of Earth Sciences, University of California, Riverside, CA 92521, USA}
\affiliation{NASA Astrobiology Institute -- Alternative Earths Team, University of California, Riverside, CA 92521, USA}
\affiliation{Blue Marble Space Institute of Science, 1001 4th Ave, Suite 3201, Seattle, WA 98154, USA}

\author{Yuka Fujii}
\affiliation{Earth-Life Science Institute, Tokyo Institute of Technology, Tokyo, 152-8550, Japan}
\affiliation{NASA Goddard Institute for Space Studies, New York, NY 10025, USA}

\author{Rodrigo Luger}
\affiliation{Center for Computational Astrophysics, Flatiron Institute, New York, NY 10010, USA}
\affiliation{NASA Astrobiology Institute -- Virtual Planetary Laboratory Team, USA}

\author{Tyler D. Robinson}
\affiliation{NASA Astrobiology Institute -- Virtual Planetary Laboratory Team, USA}
\affiliation{Department of Physics \& Astronomy, Northern Arizona University, Flagstaff, AZ 86011, USA}

\begin{abstract}

Rotational mapping and specular reflection (glint) are two proposed methods to directly detect liquid water on the surface of habitable exoplanets. However, false positives for both methods may prevent the unambiguous detection of exoplanet oceans. We use simulations of Earth as an exoplanet to introduce a combination of multiwavelength, multiphase, time-series direct-imaging observations and accompanying analyses that may improve the robustness of exoplanet ocean detection by spatially mapping the ocean glint signal. As the planet rotates, the glint spot appears to ``blink'' as Lambertian scattering continents interrupt the specular reflection from the ocean. This manifests itself as a strong source of periodic variability in crescent-phase disk-integrated reflected light curves. We invert these light curves to constrain the longitudinal slice maps and apparent albedo of multiple surfaces at both quadrature and crescent phase. At crescent phase, the retrieved apparent albedo of ocean-bearing longitudinal slices is increased by a factor of 5, compared to the albedo at quadrature phase, due to the contribution from glint. The land-bearing slices exhibit no significant change in apparent albedo with phase. The presence of forward-scattering clouds in our simulated observation increases the overall reflectivity toward crescent, but we find that clouds do not correlate with any specific surfaces, thereby allowing for the phase-dependent glint effect to be interpreted as distinct from cloud scattering. Retrieving the same longitudinal map at quadrature and crescent phases may be used to tie changes in the apparent albedo with phase back to specific geographic surfaces (or longstanding atmospheric features), although this requires ideal geometries. We estimate that crescent-phase time-dependent glint measurements are feasible for between 1 and 10 habitable zone exoplanets orbiting the nearest G, K, and M dwarfs using a space-based, high-contrast, direct-imaging telescope with a diameter between 6 and 15 m.

\vspace{3em}

\end{abstract}

\section{Introduction} \label{sec:intro}

Finding and characterizing habitable exoplanets is an emerging frontier, with the detection of liquid surface water as a fundamental goal on the path towards discovering life beyond the Solar System. A promising few small exoplanets have been discovered in the habitable zones of their parent stars \citep{Anglada-Escude2016, Gillon2016, Gillon2017, Dittmann2017}. 
These potentially habitable exoplanets are likely high priorities for upcoming observations to characterize their environments and search for signs of habitability.
However, characterizing the nature of a planet's habitability will likely require many different lines of evidence synthesized to assess the likelihood of surface liquid water \citep[for a recent review, see][]{Robinson2018, Meadows2018c}.

Detecting an ocean on the surface of an exoplanet is the most straightforward way to confirm habitability, and high-contrast direct imaging is the most promising type of observation to accomplish this. 
Unlike transmission spectroscopy, which cannot probe the surface and near-surface atmosphere \citep{GarciaMunoz2012, Betremieux2014, Misra2014}, direct imaging can probe all the way to the planetary surface, and is less susceptible to obscuration by atmospheric extinction \citep{Fortney2005}, such as by aerosols \citep{Kreidberg2014}. As a result, only direct imaging offers a feasible path to exoplanet ocean discovery. 
There are two ways exoplanet oceans could be directly detected using direct imaging of planetary reflected light: spatial mapping using the planet's rotational lightcurves, and ocean glint using phase-dependent and/or polarization observations \citep{Zugger2010}. 

Although any given snapshot of a directly-imaged exoplanet is disk-integrated and contains no spatial information, as the planet rotates on its axis, time-series observations permit spatial mapping of  heterogeneities in the atmosphere and on the surface \citep{Ford2001, Palle2008, Oakley2009}. Multi-wavelength, time-series observations can be used to infer the longitudinal map and color of multiple surfaces on a planet \citep{Cowan2009, Cowan2013a}, and multi-epoch observations can leverage the planet's obliquity and/or inclination to map latitudes and construct two-dimensional color surface maps of exoplanets \citep{Kawahara2010, Kawahara2011, Fujii2012, Berdyugina2017}. Such color maps could be used to identify large oceans on exoplanets. 

An alternative approach to detecting exoplanet oceans is to observe glint---the specular reflection typical of a flat surface or liquid when viewed at an oblique angle relative to the illumination source---a remotely detectable characteristic of the Earth ocean \citep[][]{Sagan1993}. Glint has been suggested as a promising habitability marker \citep{Williams2008, Robinson2010}, and produces a substantial relative increase in the brightness of Earth when viewed at crescent phases \citep{Robinson2014}. Targeted crescent phase observations could be used to identify the reflected light glint signature of an ocean-bearing exoplanet.

Although spatial mapping and glint are promising, both have degeneracies in interpretation that may limit the confidence with which each technique may be used to directly detect oceans. Lightcurve inversion techniques to identify the albedo and geography of multiple surfaces, including oceans \citep[e.g.][]{Cowan2009}, are under-constrained inference problems that suffer from mathematical degeneracies \citep{Fujii2017}, and have traditionally required \textit{a priori} assumptions about the surface composition to find unique solutions \citep[e.g.][]{Fujii2010}. More recently, \citet{Fujii2017} imposed the simple physical prior that the surface albedo be bounded between zero and one. This was insufficient to fully break this ``rotational unmixing'' degeneracy, but showed promise for achieving simultaneous constraints on both the color and geography of Earth-like oceans due to their low albedo and vast global distribution. 
Furthermore, simply inferring a dark blue surface is not a definitive ocean detection, as Rayleigh scattering above a grey surface can masquerade as ``blue'' in color \citep[][]{Krissansen-Totton2016}. Glint also has potential false positives.  The non-Lambertian increase in reflectivity towards crescent phase that is characteristic of ocean glint can occur independent of a glinting ocean due to forward scattering from clouds and hazes, as observed for Venus \citep[e.g.][]{Mallama2006} and Earth \citep[e.g.][]{Robinson2010}. Furthermore, high latitudes, which are more-likely to be ice-covered, have an increased contribution to the disk-integrated flux at crescent phase relative to more illuminated phases, which can naturally cause an increase in reflectivity towards crescent phase  \citep{Cowan2012}. Additional strategies and techniques are needed to help increase the robustness of ocean detection. 

Here we introduce a sequence of direct imaging observations that combine longitudinal mapping with phase-dependent glint measurements to increase the robustness of ocean detection. Specifically, multi-wavelength, time-series observations at multiple phase angles will enable spectral unmixing as a function of illumination. An exoplanet in a nearly edge-on inclination orbit, with a negligible obliquity will have rotational lightcurves that probe the same latitudes at all orbital phases, and therefore the two-dimensional mapping problem \citep[e.g.][]{Fujii2012} will be intractable as longitudinal maps will appear the same at all observed phases. However, recognizing the same longitudinal map at another phase angle would offer a second look at the same surface, thereby allowing observers to trace the reflective properties of a given geographical surface as a function of illumination angle. Thus, the reoccurring map would lend confidence to any assumption that the same surface is being viewed, while the surface albedo necessary to reproduce the observed lightcurves would reveal its scattering phase function. In this way, it may be possible to infer that the same surface that appears dark blue at gibbous phases is significantly more reflective at crescent phase, thereby proving the non-isotropic scattering behaviour of the surface. 

In this paper we demonstrate the potential utility of multi-phase longitudinal mapping as it may be applied to detect oceans on directly-imaged exoplanets. In \S\ref{sec:methods} we describe our methods used to generate synthetic lightcurves of Earth and to invert them to map the planet. In \S\ref{sec:results} we present results on multi-phase longitudinal mapping and observational feasibility tests. We offer a discussion in \S\ref{sec:discussion} and conclude in \S\ref{sec:conclusion}.

\section{Methods} \label{sec:methods}

To investigate and develop a more robust technique for exoplanet ocean detection, we forward model multi-wavelength, time-series, photometry of the disk-integrated Earth, apply realistic noise for direct-imaging exoplanet observations, and then invert the problem to solve for the geographical maps and colors of dominant surface types. The methods to generate synthetic lightcurves of Earth and fit maps to the lightcurves are described in the following subsections.

\subsection{Generating Synthetic Lightcurves}

We use the VPL Earth Model to create high spectral resolution, time-dependent, disk-averaged observations of Earth at multiple phases \citep{Robinson2011}. The VPL Earth Model gathers observed cloud and ice coverage, surface wind speed, and atmospheric thermal structure and gas mixing ratio profiles from Earth observing satellites. These Earth science data products are used as inputs to the VPL Earth Model, which maps them to a HEALPix\footnote{http://healpix.sf.net/} \citep{Gorski2005} projection of Earth. 
Multi-stream, multi-scattering, line-by-line radiative transfer calculations are then performed across the grid of cloud coverages and illumination angles, for each HEALPix pixel, using the Spectral Mapping and Atmospheric Radiative Transfer (SMART) model \citep[][developed by D. Crisp]{Meadows1996}. SMART, and by extension the VPL Earth Model, include the spectral reflection of the Sun/star off the ocean using the Cox-Munk glint model \citep{Cox1954}. Performing SMART calculations yields a database of radiances that contain contributions emerging from different regions of Earth at any given time. Spatially- and spectrally-resolved data cubes of Earth are constructed according to user specified sub-solar and sub-observer longitudes and latitudes \citep{Robinson2011}. An example VPL Earth model image of the crescent-phase Earth with a clearly visible glint spot is provided in \href{http://iopscience.iop.org/article/10.1088/2041-8205/721/1/L67/meta\#apjl366473f1}{Figure 1} of \citet{Robinson2010}.
These data cubes are then integrated over the visible disk to give a single composite disk-integrated spectrum.

The VPL Earth Model has been validated against spacecraft observations of Earth, including spatially resolved and time-dependent observations made by the EPOXI mission \citep{Robinson2011}, and \textit{phase-dependent} measurements of Earth showing the glint effect, that were taken by the Lunar Crater Observing and Sensing Satellite (LCROSS) \citep{Robinson2014}. 
We refer the reader to \citet{Robinson2011, Robinson2014} and \citet{Schwieterman2015, Schwieterman2016}\footnote{Earth spectral database used for this work is available upon request. Previous similar datasets are available at: http://depts.washington.edu/naivpl/content/spectral-databases-and-tools} for a more complete description of the VPL Earth Model and its validation. 

The VPL Earth Model is used to generate multiple different Earth and Earth-like spectral models that enable longitudinal mapping investigations at multiple phases. We make a simplifying assumption that the sub-solar and sub-observer latitudes are fixed at the equator to simulate a planet with zero obliquity in an edge-on orbit. We compute time-dependent spectra at three phase angles: $45^{\circ}$, $90^{\circ}$, and $135^{\circ}$, which we refer to as gibbous, quadrature, and crescent, hereafter. At each phase we simulate 120 consecutive hours (5 days) of high-resolution, disk-integrated spectra at a 1 hour cadence. We neglect the small (${<} 5^{\circ}$) phase change of the planet over the 120 hours of Earth modeling. 
The computational expense of the VPL Earth Model places limits on the observational duration and observational cadence that we may consider in this work, and in \S\ref{sec:results:mapping}, we focus on experiments with 100 hours of data collection at a 1 hour cadence.
We simulate Earth during northern spring. Datasets are generated both with true time-varying cloud coverage, and without clouds. 

To simulate realistic observations of our VPL Earth Model spectra as an analog for a directly-imaged exoplanet, we use an open-source Python coronagraph noise model\footnote{https://github.com/jlustigy/coronagraph}, originally developed by \citet{Robinson2016} for the WFIRST exoplanet direct imaging mission and expanded upon in subsequent studies \citep{Luger2017, Meadows2018}. The model simulates planetary and background photon count rates as a function of wavelength using analytic parameterizations for telescope, instrumental, and astrophysical noise sources. These sources include Poisson noise, read noise and dark current, coronagraph speckles, scattering off of zodiacal and exo-zodiacal dust, and thermal emission from the mirror. The telescope is assumed to perform a roll maneuver to allow for background subtraction \citep[c.f.][]{Brown2005}, which increases the background by a factor of two \citep{Robinson2016}.  

We perform coronagraph simulations for various telescope designs that encompass telescope mission concepts currently under study by NASA Science \& Technology Definition Teams (STDTs) in advance of the 2020 decadal survey. Many of the telescope, instrument, and system parameters used here are the same as those used in \citet[][see their Table 2 and Table 3]{Robinson2016}, with the exceptions listed in Table \ref{tab:params}. We adopt a raw contrast of $C = 10^{-10}$ to pursue Earth-like exoplanets orbiting Sun-like stars. We consider telescope diameters between 4 - 15 meters, and end-to-end throughputs between 0.02 - 0.5. These ranges generally encapsulate those relevant to the Large UV/Optical/IR \citep[LUVOIR\footnote{https://asd.gsfc.nasa.gov/luvoir/}; e.g.][]{Dalcanton2015, Bolcar2017} Surveyor and Habitable Exoplanet Imaging Mission \citep[HabEx\footnote{http://www.jpl.nasa.gov/habex/}; e.g.][]{Mennesson2016} telescope mission concepts. We adopt a highly optimistic inner working angle (IWA) of $1 \lambda/D$ and outer working angle (OWA) of $60 \lambda/D$ so that our simulated lightcurves solely represent the photon counting statistics. However, we separately consider constraints on the IWA in \S\ref{sec:results:iwa}. We note that, like \citet{Robinson2016}, the IWA and OWA are modeled as strict cutoffs---a step function in throughput from 100\% to 0\%---whereas formally they are defined as the 50\% off-axis throughput point. As a result, our coronagraph modeling neglects throughput effects as the IWA and OWA are approached and exceeded. 

\begin{deluxetable}{llcl}
\tablewidth{0.47\textwidth}
\tabletypesize{\scriptsize}
\tablecaption{Telescope, instrument, and system model parameters\label{tab:params}}
\tablehead{\colhead{Parameter} & \colhead{Description} & \colhead{Value(s)}} 
\startdata
$d$  & distance to system  & $5$ pc \\
$R_p$ & planetary radius & $1 \ R_{\oplus}$ \\
$\alpha$ & planet phase angle & $45^{\circ}, 90^{\circ}, 135^{\circ}$ \\
$C$   & coronagraph design contrast & $10^{-10}$ \\
$D_{e-}$ & dark current & $10^{-4}$ s$^{-1}$ \\
$D$ & telescope diameter & $(4, 15)$ m \\
$\mathcal{R}$ & instrument spectral resolution & R$200$ \\
$\mathcal{T}$ & end-to-end throughput & $(0.02, 0.5)$ \\
$\theta_{\text{IWA}}$ & coronagraph inner working angle & $ > 1 \lambda/D$ \\
$\theta_{\text{OWA}}$ & coronagraph outer working angle & $ < 60 \lambda/D$
\enddata
\tablecomments{See \citet{Robinson2016} for a complete list of coronagraph parameters and their nominal values. Comma separated values are all considered and comma separated values in parentheses represent explored ranges.}
\end{deluxetable}

We construct simultaneous, multi-wavelength lightcurves by simulating coronagraph spectroscopy with a resolving power of $\mathcal{R} = \lambda / \Delta \lambda = 200$ at a one hour cadence over a modest 18\% bandpass. This gives a low signal-to-noise spectrum each hour, and from hour-to-hour the spectrophotometry samples Earth's rotational variability. While multiple hours of spectra can be co-added into a higher signal-to-noise spectrum that may be used to retrieve atmospheric gas abundances \citep{Feng2018}, here we instead co-add neighboring spectral resolution elements to increase the signal-to-noise on each hour of photometry. Since current coronagraph designs can only null over relatively modest spectral bandpasses (typically $\Delta \lambda / \lambda \approx 10 \% - 20 \% $) in a single pointing, we focused on synthetic observations in an 18\% bandwidth wavelength range between 0.67 and 0.8 $\mu$m.  In this wavelength range the Earth exhibits large amplitude rotational variability owing to surface features, such as ocean, soil, and vegetation.  This sensitivity is due in large part to the Earth's atmosphere being optically thin, with very few molecular absorption features and reduced Rayleigh scattering, at most of these wavelengths.  The one major exception is the opacity due to the 0.76 $\mu$m O$_2$ absorption band, which is likely to be one of the most sought-after atmospheric diagnostics for Earth-like planets \citep{Meadows2017, Meadows2018b}. Consequently, the same multi-wavelength, time-resolved data that is used to map the planet and search for oceans can be coadded to characterize the planet's atmosphere and search for signs of life. We bin the spectrophotometry into two photometric bands, centered at $\lambda_1 = 0.70$ $\mu$m and $\lambda_2 = 0.77$ $\mu$m with widths $\Delta \lambda_1 = 0.08$ $\mu$m and $\Delta \lambda_2 = 0.07$ $\mu$m, sampled over many consecutive hours and with corresponding noise applied, which yield lightcurves for studying the time-variability of the exoplanet. We assume no read time, no reset time, and no down time between consecutive exposures thereby simulating a 100\% efficient duty cycle.

\subsection{Lightcurve Fitting}

Time-resolved, multi-wavelength planetary lightcurves can be used to constrain the longitudinal map of a planetary surface, using techniques of ``spectral-rotational unmixing'', which solve for the longitudinal map of $N$ different unique surface types and their geometric albedo spectrum \citep{Cowan2009, Cowan2012, Fujii2017}. 
As described in detail in \citet{Fujii2017} and summarized here for completeness, the observed apparent albedo of the planet, $A_{ij}$, at wavelength $j$ and at time $i$ can be decomposed into
\begin{equation}
\label{eqn:fitlc}
A_{ij} = \sum_{i,k} \tilde{f}_{ik} a_{kj}
\end{equation}
where $\tilde{f}_{ik}$ is the apparent covering fraction of the $k$-th surface at the $i$-th time and $a_{kj}$ is the apparent albedo spectrum of the $k$-th surface at the $j$-th wavelength. 

Similarly, due to the relationship between time variability and surface heterogeneity for a rotating planet, the observed apparent albedo can be written in terms of a longitudinal map. In this case, $\tilde{f}_{ik}$ may be approximated by 
\begin{equation}
\label{eqn:flk}
\tilde{f}_{ik} \approx \sum_{l} W_{il}f_{lk}
\end{equation}
where $f_{lk}$ is the area fraction of the $k$-th surface in the $l$-th longitudinal slice and $W_{il}$ is the weight of the $l$-th longitudinal slice at the $i$-th time, which is a function of the observational geometry. The approximation in Equation \ref{eqn:flk} is only strictly valid when the true area fraction within each longitudinal slice does not change or changes little across the slice for all surfaces. Plugging Equation \ref{eqn:flk} into Equation \ref{eqn:fitlc}, we reach the following expression in terms of longitudinal slices:
\begin{equation}
\label{eqn:fitmap}
A_{ij} \approx \sum_{l,k} W_{il} f_{lk} a_{kj}
\end{equation}

Both Equations \ref{eqn:fitlc} and \ref{eqn:fitmap} form the two modular forward models in the Surface Albedo Mapping Using RotAtional Inference (\samurai\footnote{https://github.com/jlustigy/samurai}) Python code developed in this work. \samurai\ attempts to solve the inverse problem for the map ($\tilde{f}_{ik}$ or $f_{lk}$) and albedo ($a_{kj}$) of $N$ different surfaces using multi-band $\{j\}$, time-series $\{i\}$ photometric observations $y_i(\lambda_j)$ and their observational uncertainties $\sigma_i(\lambda_j)$, subject to the following constraints on the model $A_{ij}$: 
\begin{subnumcases}
{}
\tilde{f}_{ik}, f_{lk} \ge 0 & for any $i$, $l$, $k$ \label{eqn:f_range} \\
\sum_{k} \tilde{f}_{ik} = \sum_{k} f_{lk} = 1 & for any $i$, $l$ \label{eqn:f_sum} \\
0 \le a_{kj} \le 1 & for any $k$, $j$ \label{eqn:a_range} 
\end{subnumcases}
Equation \ref{eqn:f_range} ensures that the area fraction of any surface $k$ at any time $i$ or in any longitudinal slice $l$ is not negative, while Equation \ref{eqn:f_sum} enforces that the area fractions for all surface types at any time or in any longitudinal slice sum to unity. Equation \ref{eqn:a_range} imposes that the albedo of a surface $k$ at any wavelength $j$ does not exceed 1. 

The hard boundaries in \ref{eqn:f_range}, \ref{eqn:f_sum}, and \ref{eqn:a_range} make it difficult to efficiently sample the likelihood space using Markov Chain Monte Carlo (MCMC). One way around this is to reparametrize so that the domain of our parameters extends from $-\infty$ to $+\infty$. This can be done by introducing the transformed parameters
\begin{subnumcases}
{}
\label{eqn:b_kj}
b_{kj} = \ln \left ( \frac{a_{kj}}{1 - a_{kj}} \right ) & for any $k$, $j$ \label{eqn:trf1} \\
\label{eqn:g_lk}
g_{lk} = \ln \left ( \frac{f_{lk}}{1 - \sum_{m = 1}^{k} f_{lm}} \right ) & for $k = 1, ..., N-1$ \label{eqn:trf2}
\end{subnumcases}
where $N$ is the number of surface types included in the model. Note that conditions \ref{eqn:f_range}, \ref{eqn:f_sum}, and \ref{eqn:a_range} are implicitly satisfied by this reparameterization, and there is a one-to-one correspondence between albedo matrices $a$ and $b$, and between area covering fraction matrices $f$ and $g$. Similar reparameterizations have been applied to gas mixing ratios in exoplanet atmospheric retrieval problems \citep[c.f.][]{Benneke2012}. 

It is important to note that when reparameterizing our model, we must also transform the prior probability of each model parameter. Since we have no \emph{a priori} information on the albedos and area fractions of individual surfaces, we enforce a flat prior for these variables. This can be obtained by imposing the prior probabilities 
\begin{align}
    \label{eqn:prior_alb}
    \mathcal{P}(b_{kj}(a_{kj})) & \propto \frac{e^{b_{kj}}}{(1 + e^{b_{kj}})^2} \\ 
    \label{eqn:prior_area}
    \mathcal{P}(g_{lk}(f_{lk})) & \propto \frac{e^{g_{lk}}}{(1 + e^{g_{lk}})^2}
\end{align}
on our transformed variables $b$ and $g$, where $\mathcal{P}$ is a probability distribution function.
Note that Equation \ref{eqn:prior_area} does not contain the covariance term that arises due to the sum in Equation \ref{eqn:g_lk}. As a result, Equation \ref{eqn:prior_area} does not guarantee a flat prior for three or more surfaces, and thus all investigations in this paper are performed with only two surfaces. 
We impose one additional prior $\mathcal{O}(g_{lk})$ to ensure that surface covering fractions are ordered from largest to smallest to break a surface labeling degeneracy. Considering all elements in $b$ and $g$, and $\mathcal{O}$, our full prior is
\begin{align}
\mathcal{\pi}(b_{kj}, g_{lk}) &= \mathcal{P}(b_{kj}) \mathcal{P}(g_{lk}) \mathcal{O}(g_{lk}) \\ 
&\propto \left ( \prod_{k,j} \frac{e^{b_{kj}}}{(1 + e^{b_{kj}})^2} \right ) \left ( \prod_{l,k} \frac{e^{g_{lk}}}{(1 + e^{g_{lk}})^2} \right ) \mathcal{O}(g_{lk}) .
\end{align}

The aim of our parameter estimation is to evaluate the probability of the albedo and map parameters---the posterior probability distribution function (PDF)---given multi-wavelength, time-series data. By Bayes' Theorem, the posterior PDF is  
\begin{equation}
\label{eqn:lnp}
    \mathcal{P}(b_{kj}, g_{lk} | y_{ij}) \propto \mathcal{L}(y_{ij} | b_{kj}, g_{lk}) \times \mathcal{\pi}(b_{kj}, g_{lk})
\end{equation}
where the likelihood, $\mathcal{L}(y_{ij} | b_{kj}, g_{lk})$ is easily calculated from the model and data,
\begin{equation}
    \ln \mathcal{L}(y_{ij} | b_{kj}, g_{lk}) = - \frac{1}{2} \sum_{i,j} \left ( \frac{y_{ij} - W_{i}^{l} f_{l}^{k} a_{kj}}{\sigma_{ij}} \right )^2 .
\end{equation}

Unless otherwise stated, we sample the log-probability (log of Equation \ref{eqn:lnp}) using the \emcee\footnote{\emcee\: \url{https://github.com/dfm/emcee}} implementation of MCMC \citep{Foreman-Mackey2013}. We use a linear combination of $15\%$ affine-invariant \citep{Goodman2010}, $70\%$ differential evolution \citep{Nelson2014}, and $15\%$ differential evolution snooker \citep{Ter2008} MCMC proposal steps to improve convergence in a likelihood space that is, at times, multi-modal, with different sets of surface spectra and geography that can reproduce the observed lightcurves. We transform our sampling parameters $b$ and $g$ back to physical parameters $a$ and $f$ after the MCMC. This analysis results in posterior PDFs for the wavelength-dependent albedo and longitudinal area fraction of each surface included in the fit. Deriving PDFs for our model parameters allows us to quantify uncertainties on our inferences so that we may make robust statements about any changes seen when the experiment is repeated at another phase.

\section{Results} \label{sec:results}

We performed spectral unmixing experiments at quadrature, crescent, and gibbous phase to determine if a combination of phase-dependent surface mapping and glint observations could provide more robust constraints on the presence of an ocean for an Earth analog exoplanet. We then quantified the observational feasibility and telescope requirements for this technique with respect to coronagraph inner working angle and observed photometric signal-to-noise.  

\subsection{Multi-Phase Longitudinal Mapping}
\label{sec:results:mapping}

\begin{figure*}
\centering
\includegraphics[width=\textwidth]{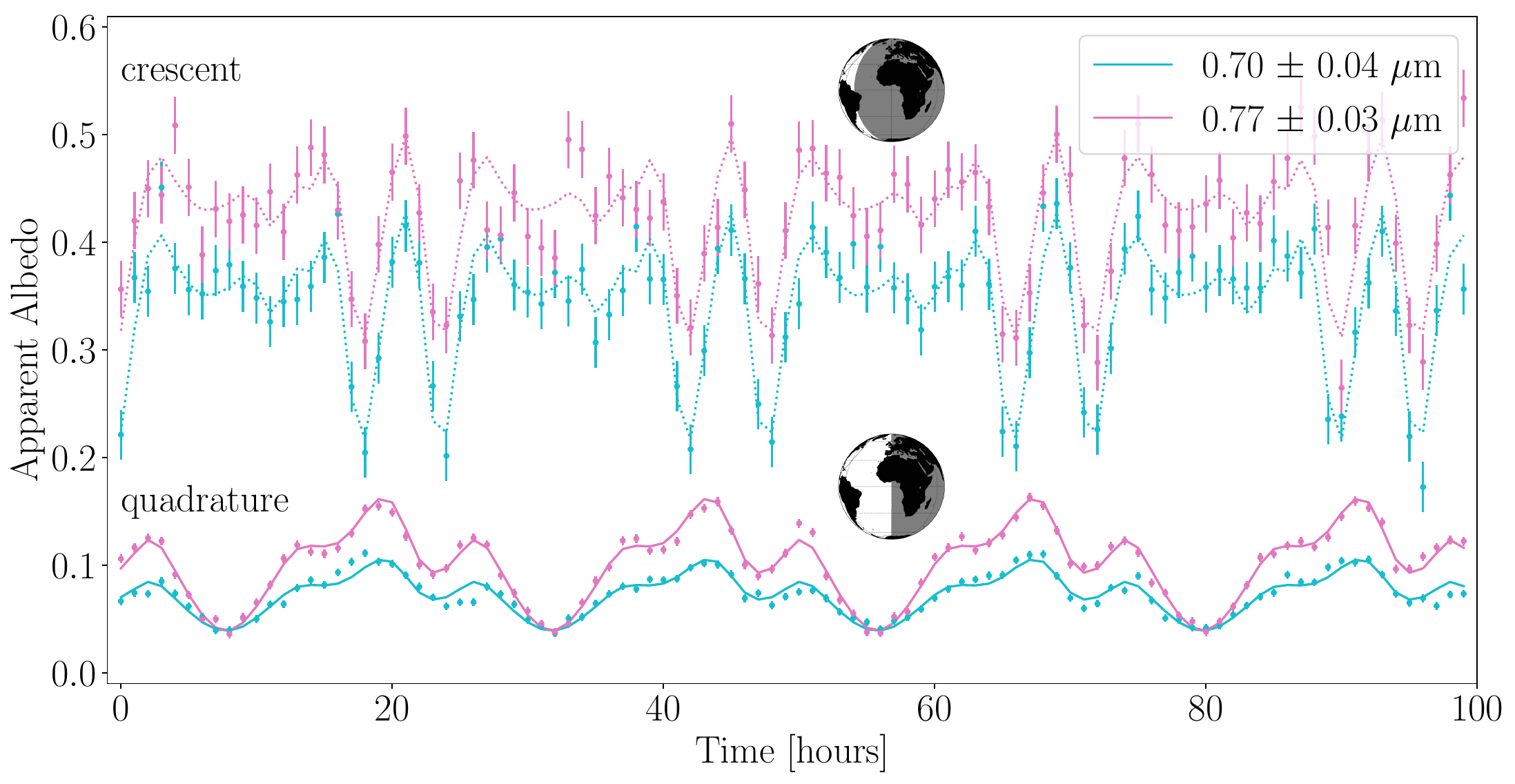}
\includegraphics[width=\textwidth]{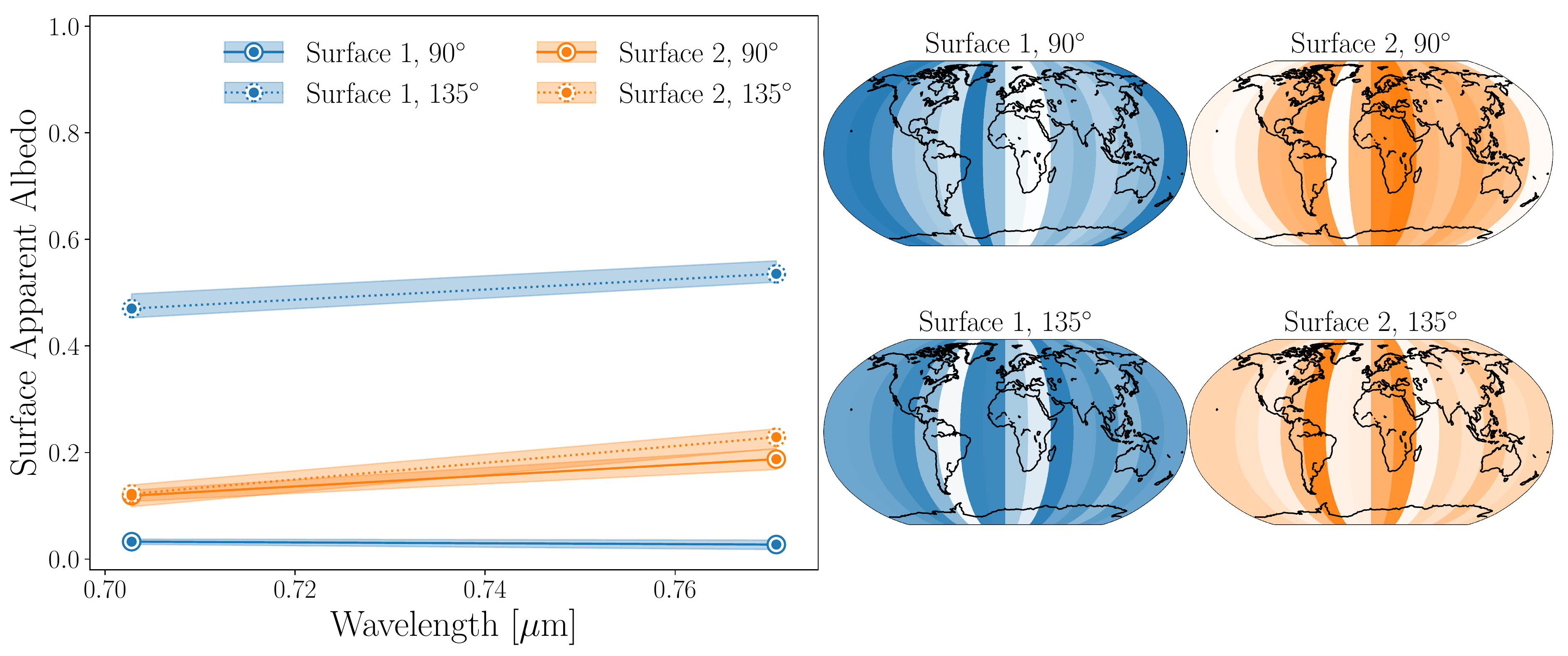}
\caption{Multi-phase inverse problem example for the cloudless Earth. \textit{Top:} Synthetic lightcurves for two wavelength intervals with error bars for the cloudless Earth, and retrieved fits to the observations (lines) at quadrature (solid lines) and crescent (dotted lines) phase. Data were simulated for the cloudless Earth at 5 pc observed with a 15 m telescope with 50\% throughput, which yielded a 1-hour photometric precision of 4.5\% at quadrature and 6.5\% at crescent. 
\textit{Bottom Left:} Retrieved surface albedos from a two surface fit to rotational lightcurves of the cloudless Earth at 90$^{\circ}$ phase and 135$^{\circ}$ phase. The color of each line identifies the surface and the line style denotes the phase. The envelope about each line shows the 1$\sigma$ interval about the median posterior PDF derived from MCMC. From quadrature to crescent phase, the average albedo of Surface 1 increases by a factor of ${\sim} 16$ with 15$\sigma$ confidence, while the average albedo of Surface 2 increase by ${\sim}15\%$ at ${<} 1 \sigma$. 
\textit{Bottom Right:} Median retrieved longitudinal slice maps. Each longitudinal slice is colored proportional to the retrieved area covering fraction of the given surface, with white meaning that none of the given surface is found within the slice. Maps are shown in the Robinson projection.}
\label{fig:cloudless_maps}
\end{figure*}

Our first multi-phase mapping experiment focused on models without clouds to provide an ideal end-member to isolate the signal from the ocean. This also allowed us to assess the unique impact of clouds once they were included.

The upper panel of Fig. \ref{fig:cloudless_maps} shows the quadrature ($\alpha = 90^{\circ}$; solid lines) and crescent ($\alpha = 135^{\circ}$; dotted lines) phase multi-wavelength lightcurves and median model fits to these data. The lightcurves were simulated for an optimistic case of Earth at 5 pc, observed for 100 hours at each phase, at a 1 hour cadence, with a 15 m, 50\% throughput space telescope. The spectrophotometry were binned into two wavelength intervals to allow for slopes in the albedo of each surface. This yielded lightcurves at $0.70 \pm 0.04$ $\mu$m and $0.77 \pm 0.03$ $\mu$m, with an average 1-hour photometric uncertainty, respectively, of 5.0\% (S/N=20.1) and 4.0\% (S/N=25.4) at quadrature, and 6.8\% (S/N=14.8) and 6.1\% (S/N=16.5) at crescent. The decrease in photometric precision seen for the crescent lightcurves in Fig. \ref{fig:cloudless_maps} is due to the decrease in the illuminated portion of the planet's disk, for observations using the same telescope. 

The lower panels of Fig. \ref{fig:cloudless_maps} show the retrieved apparent albedo and longitudinal slice maps of each surface, for a two surface, 16 slice fit to the synthetic multi-wavelength lightcurves shown in the upper panel. The choice of 16 slices was selected so that the angular slice width ($22.5^{\circ}$) divides the change in phase ($45^{\circ}$) by an integer, thereby allowing the quadrature and crescent phase maps to be shifted and aligned. The slice maps are displayed on a Robinson map projection, and use color intensity to indicate the median retrieved area covering fraction in each slice. The retrieved surface albedos are shown with a 1$\sigma$ uncertainty envelope about the median posterior PDFs derived from MCMC. 
The apparent albedo of Surface 1 (blue) exhibited a strong increase from quadrature to crescent phase, becoming ${\sim} 16 \times$ brighter, while the albedo of Surface 2 (orange) increased by only ${\sim}15\%$, which was within the derived uncertainties.   

\begin{figure}
\centering
\includegraphics[width=.47\textwidth]{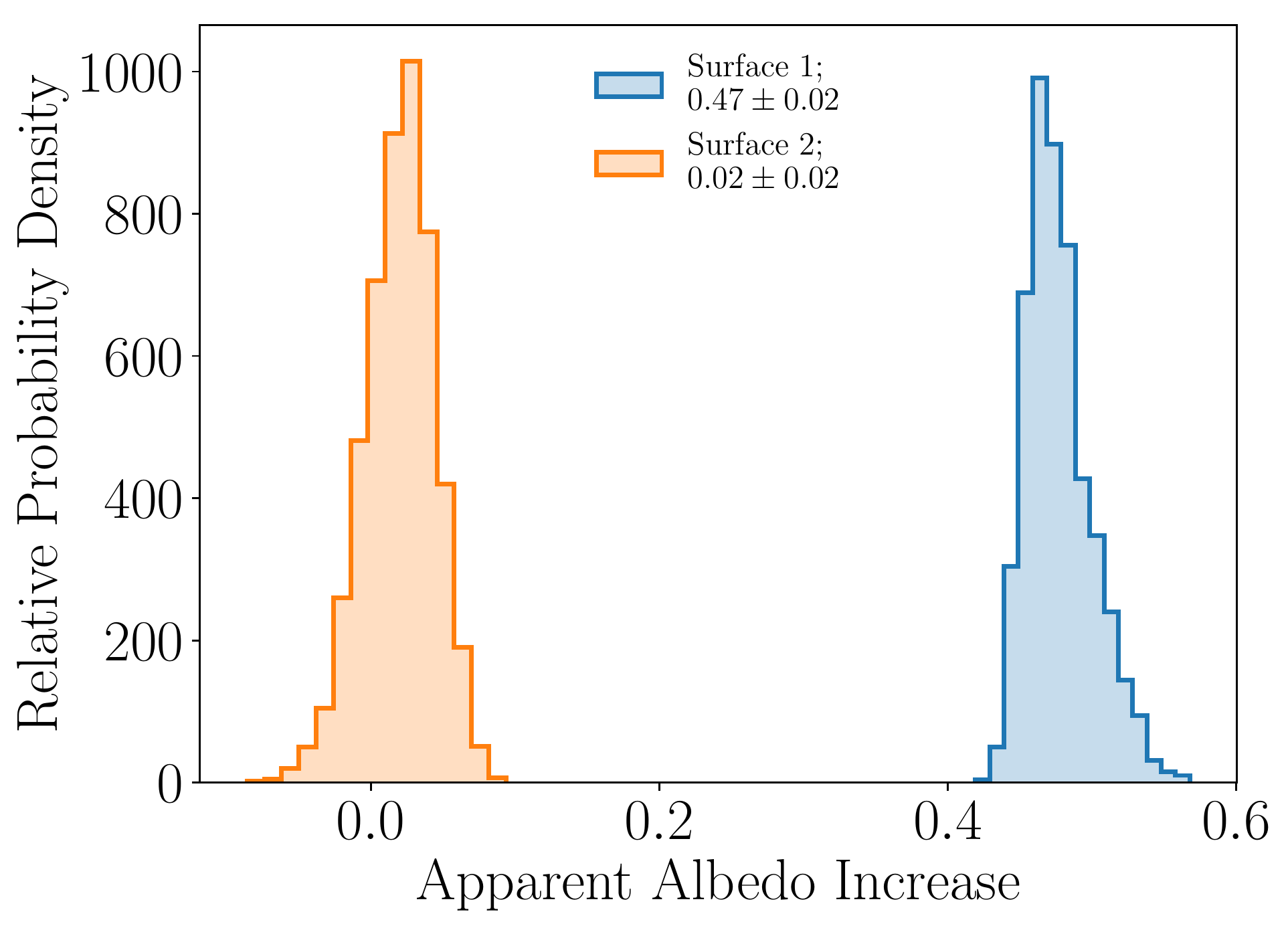}
\caption{Histograms showing the increase in albedo for Surface 1 (blue) and Surface 2 (orange) between quadrature and crescent phase observations of the cloudless Earth. Surface 1 increased in apparent albedo by $0.47 \pm 0.02$ from quadrature to crescent phase, indicating that it has a highly non-Lambertian scattering phase function. Surface 2 has a negligible albedo increase, indicating that it is consistent with Lambertian scattering.}
\label{fig:cloudless_albedo_compare}
\end{figure}

Figure \ref{fig:cloudless_albedo_compare} explores the statistical significance of the increase in apparent albedo of Surface 1 and Surface 2 between the retrievals at quadrature and crescent phase.  Histograms showing the change in apparent albedo of each surface were produced by taking 5000 random samples from the wavelength-averaged posterior PDFs at each phase and then subtracting the quadrature phase result from the crescent phase result. Surface 1's apparent albedo increased by $0.47 \pm 0.02$, while Surface 2's apparent albedo remained constant with phase ($0.02 \pm 0.02$). 

\begin{figure}
\centering
\includegraphics[width=.23\textwidth]{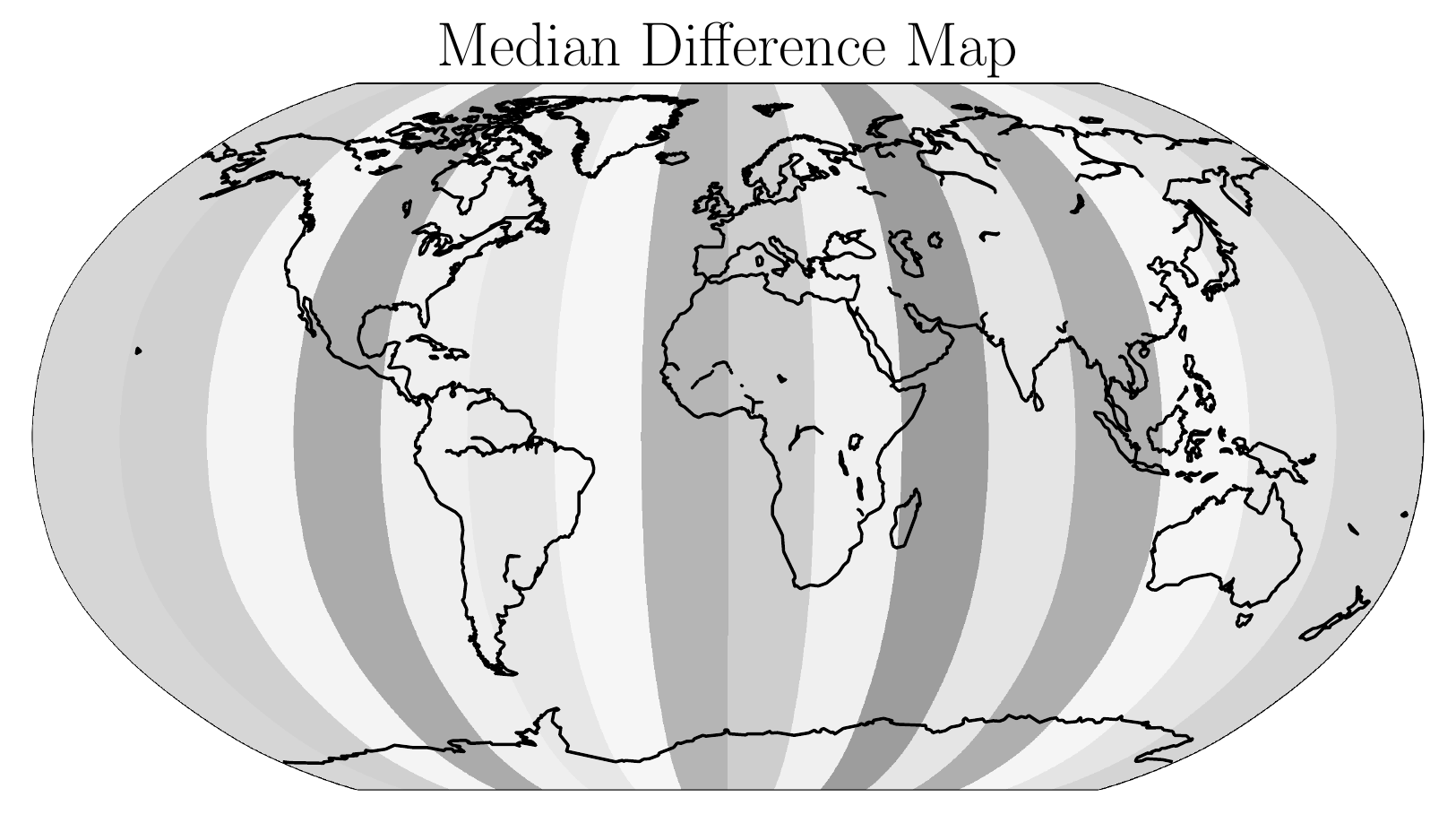}
\includegraphics[width=.23\textwidth]{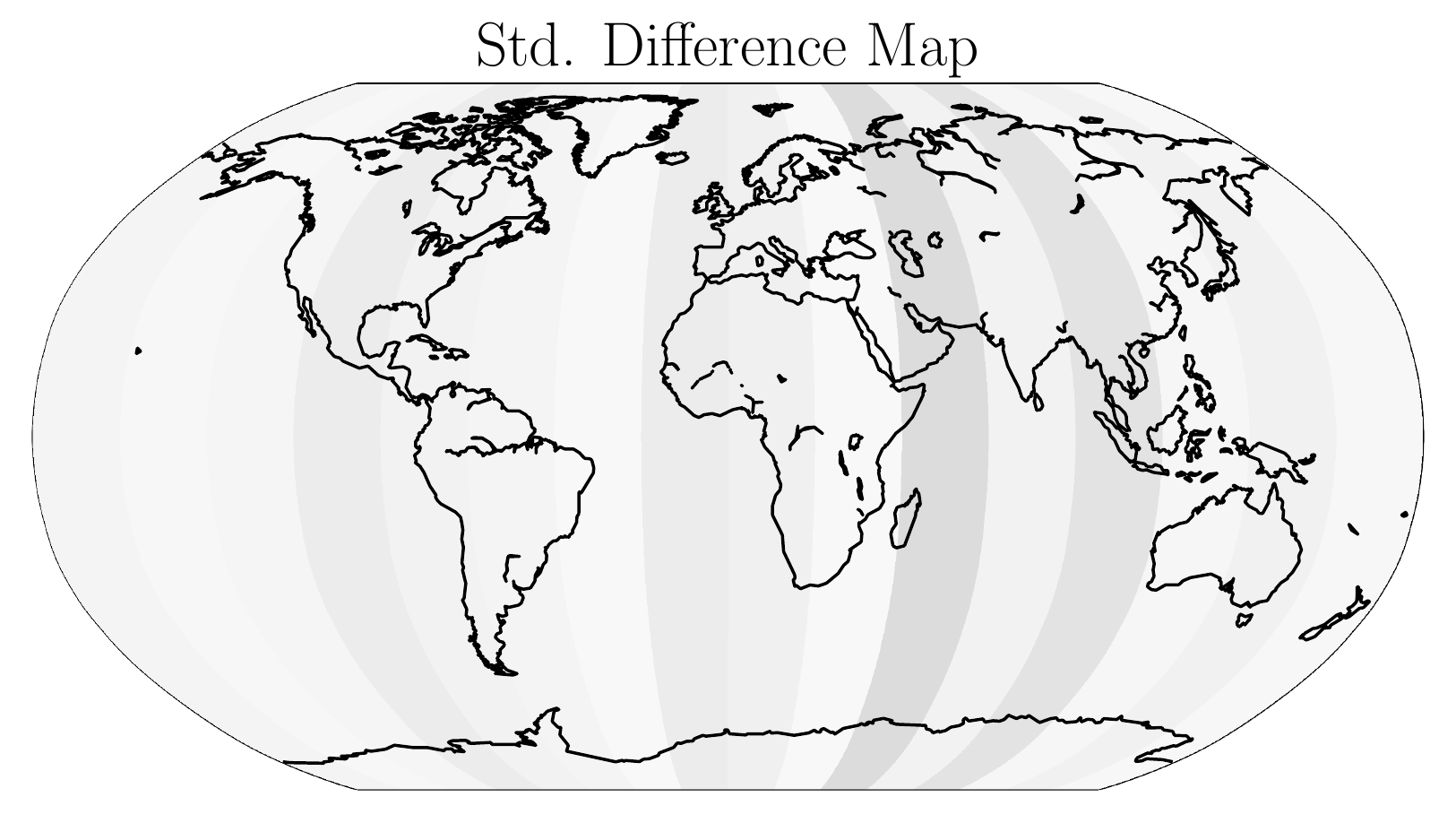}
\includegraphics[width=.47\textwidth]{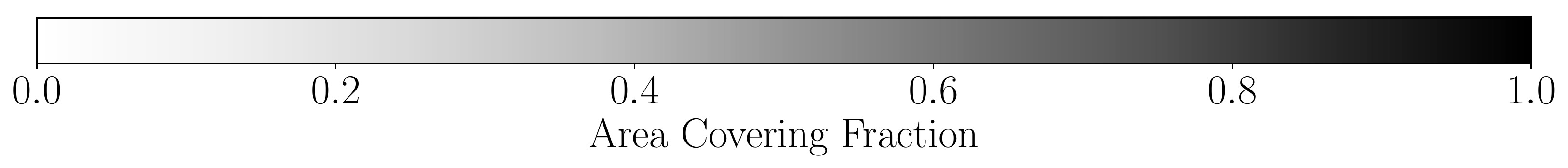}
\includegraphics[width=.47\textwidth]{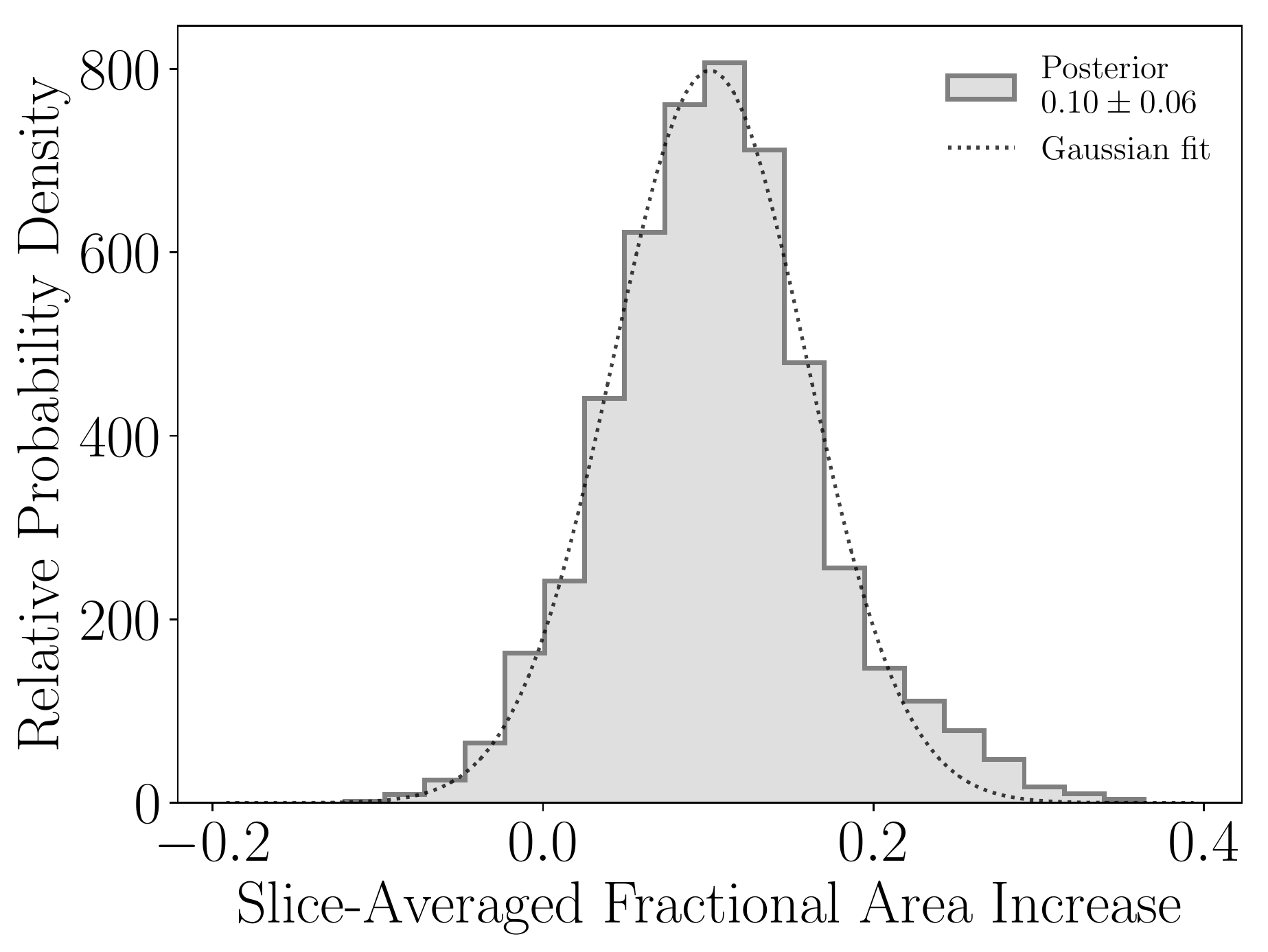}
\caption{\textit{Top:} Longitudinal slice maps showing the median (left) and standard deviation (right) of the change in area covering fraction between quadrature and crescent phase observations of the cloudless Earth. \textit{Bottom:} Longitudinal slice-averaged fractional area change of Surface 1 from quadrature to crescent phase. The maps agree best for slices containing majority open ocean, with the marginally statistically significant change in the maps with phase primarily driven by disagreements in land-bearing slices.}
\label{fig:cloudless_map_compare}
\end{figure}

Figure \ref{fig:cloudless_map_compare} examines the difference in the retrieved longitudinal maps with phase. A qualitative comparison shows that the maps at quadrature and crescent phase in Fig. \ref{fig:cloudless_maps} are very similar, but a direct comparison of the maps and their uncertainties at each phase reveals both regions of agreement and regions of disagreement. The two longitudinal maps in Fig. \ref{fig:cloudless_map_compare} show the median (left) and standard deviation (right) of the difference between the maps at each phase, in each slice. This is simply the difference between the two Surface 1 slice maps in Fig. \ref{fig:cloudless_maps}, with uncertainties propagated by randomly sampling the MCMC posterior distributions. Figure \ref{fig:cloudless_map_compare} is shown with respect to Surface 1, although for a two surface longitudinal model Surface 2 is simply one minus the map of Surface 1 (see condition \ref{eqn:f_sum}). The median difference between the maps is less than 0.5 for all slices. The histogram in the bottom panel of Fig. \ref{fig:cloudless_map_compare} shows the slice-averaged (or total) fractional area increase with the change in phase. The crescent phase results include slightly more contribution from Surface 1 ($0.10 \pm 0.06$) than the quadrature results. However, the longitudinal maps agree best for slices containing majority open ocean, with the marginally significant change in the maps with phase primarily driven by disagreements in the land-bearing slices. This suggests that even despite the ideal system geometry, telescope noise properties, and lack of clouds, the inferred maps have intrinsic differences. 

Simulations performed at gibbous phase ($45^{\circ}$) show similar results to those at quadrature phase, and are therefore not shown. At gibbous phase the recovered apparent surface albedos are consistent with the recovered albedos at quadrature, with one very dark surface and another more reflective surface that increases in albedo from about 0.1 to 0.2 over the wavelength range of interest. However, the retrieved maps at gibbous are less precise than both the quadrature and crescent phase maps despite having a higher 1-hour photometric signal-to-noise. This difference is due to the larger phase which reflects more total light to the observer, but that light is integrated over a larger portion of the planetary disk which lowers the observer's sensitivity to each individual longitude, thereby lowering the resolution of the recovered maps \citep{Fujii2017}. 

\begin{figure}
\centering
\includegraphics[width=0.47\textwidth]{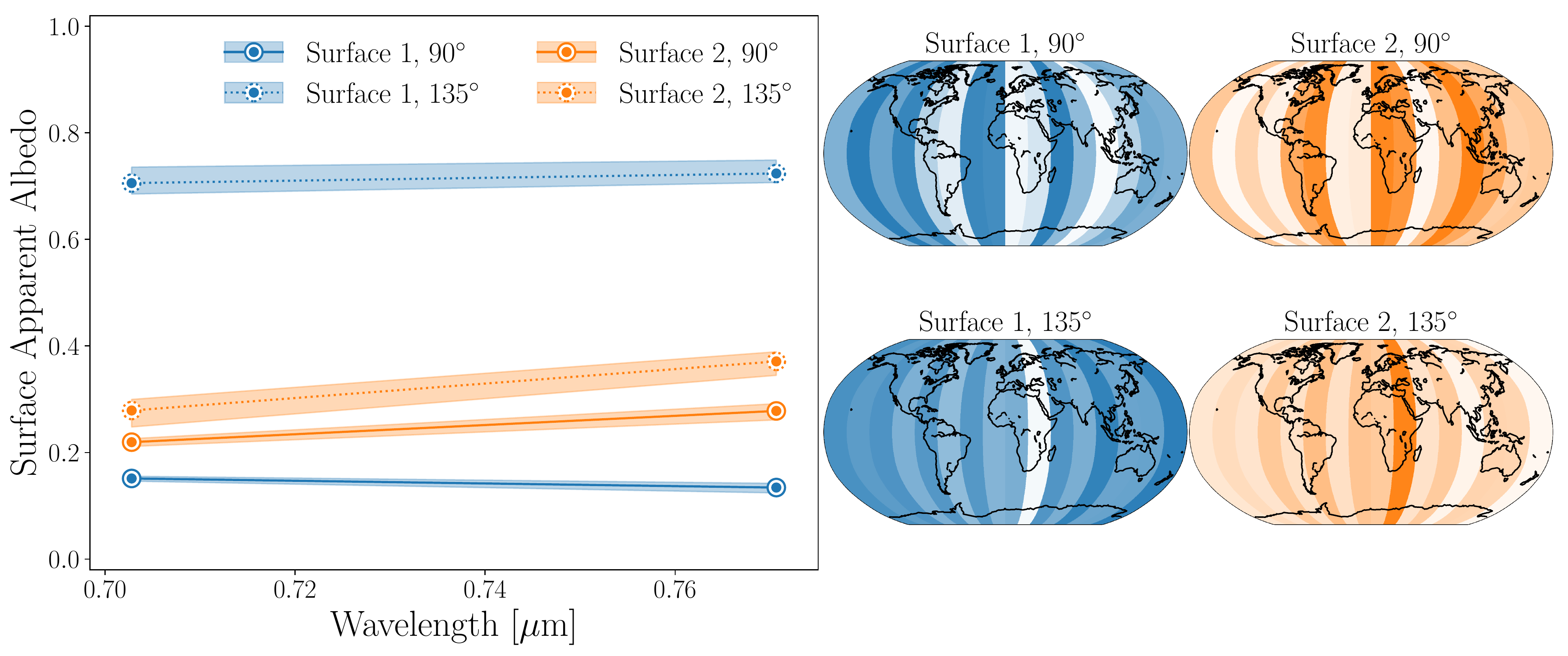}
\includegraphics[width=0.47\textwidth]{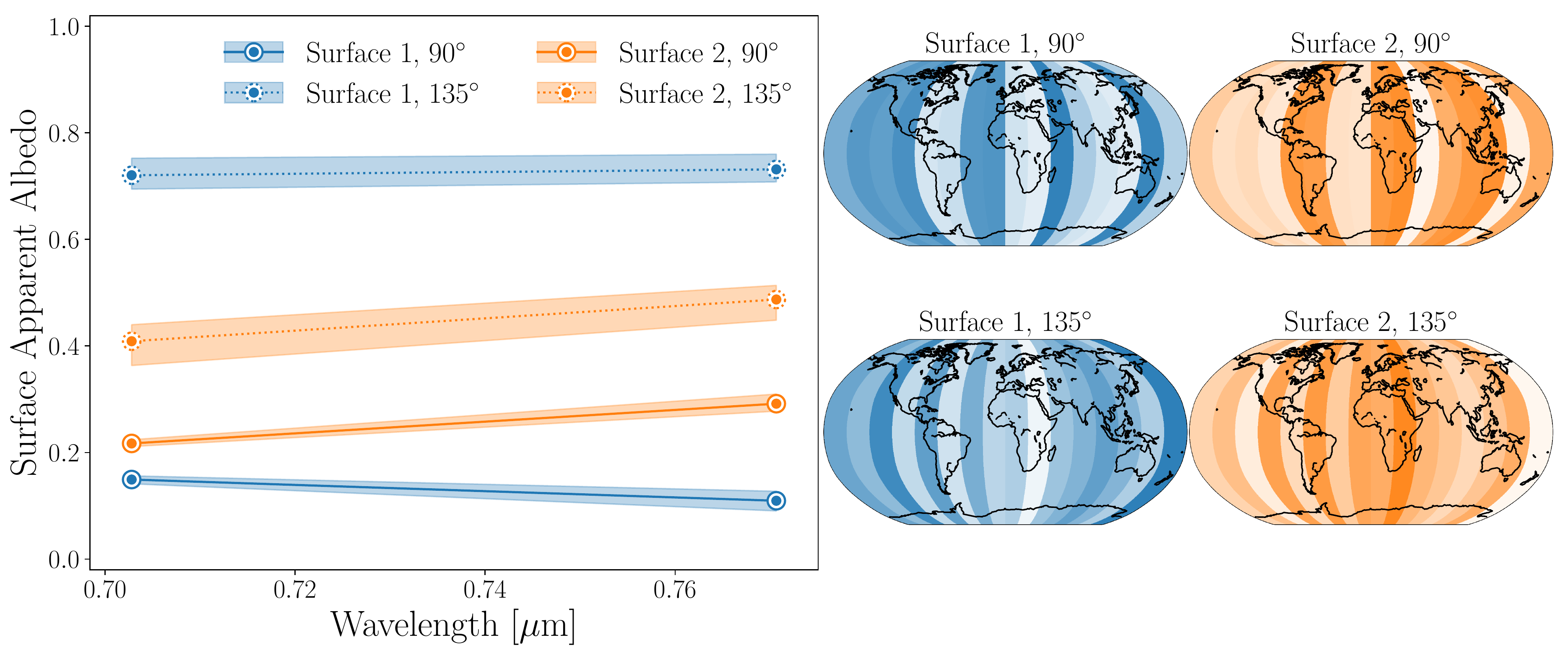}
\includegraphics[width=0.47\textwidth]{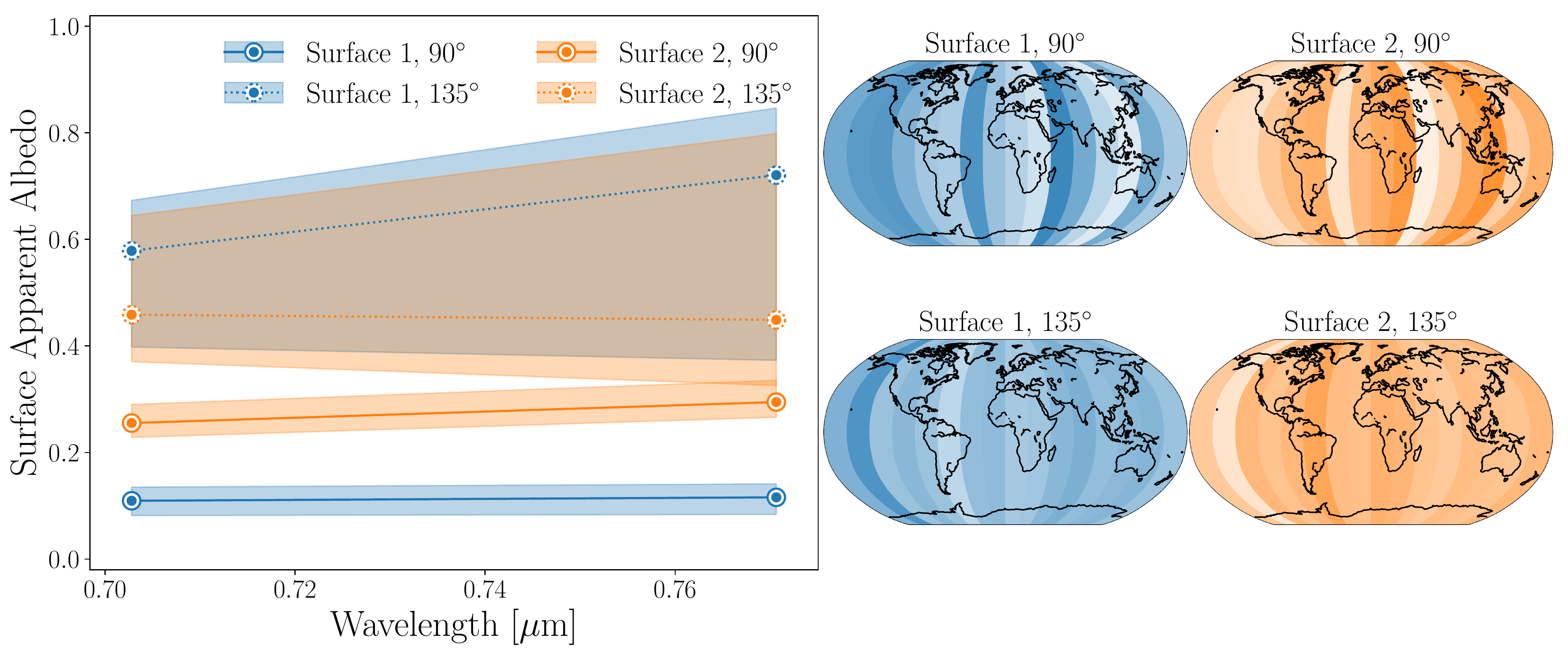}
\caption{Similar to the lower panels of Fig. \ref{fig:cloudless_maps}, but now for Earth with time-varying clouds and at three different levels of photometric precision. From top to bottom, the data have a 1-hour S/N (photometric precision) of 23 (4.3\%), 10 (10\%), and 2.4 (42\%), at quadrature, which correspond to observations of Earth at 5 pc with a 15 m and 50\% throughput telescope, a 15 m and 10\% throughput telescope, and a 6 m with 10\% throughput telescope, respectively.}
\label{fig:glint_map}
\end{figure}

We performed the same experiment on simulations of Earth with real time-varying cloud coverage. Figure \ref{fig:glint_map} shows results from multiple surface mapping retrievals performed for Earth at 90$^{\circ}$ (quadrature; solid lines) and 135$^{\circ}$ (crescent; dotted lines), where each row of plots corresponds to data generated at a different signal-to-noise ratio. The upper panel of Fig. \ref{fig:glint_map} shows longitudinal mapping results from lightcurves simulated for the optimistic case of Earth at 5 pc, observed for 100 hours at each phase, at a 1 hour cadence, with a 15 m, 50\% throughput space telescope (same setup at Fig. \ref{fig:cloudless_maps}). The middle and lower panels of Fig. \ref{fig:glint_map} show similar results to the upper panel, except they correspond to data simulated for a 15 m with 10\% throughput telescope, and a 6 m with 10\% throughput telescope, respectively. These three datasets have 1-hour photometric signal-to-noise at quadrature phase of 23 (4.3\%), 10 (10\%), and 2.4 (42\%), from the top to bottom panels. As in the cloudless case, the spectrophotometry were binned into two wavelength intervals, and the maps were fit with a two-surface, 16 longitudinal slice model. The envelope about each line in Fig. \ref{fig:glint_map} shows the 1$\sigma$ uncertainty about the median retrieved albedo spectrum parameters derived from MCMC. 

Comparing results at each phase for the highest S/N data we find that the albedo of Surface 1 exhibits the same large increase at crescent as seen for the cloudless case, even in the presence of realistic clouds, and the longitudinal maps have reasonably good agreement, particularly at longitudes between $-180^{\circ}$ and $0^{\circ}$. 
By the same test shown in Fig. \ref{fig:cloudless_albedo_compare}, Surface 1's apparent albedo increased by $0.57 \pm 0.02$ with the change in phase, while Surface 2's apparent albedo increased by a slight margin of $0.08 \pm 0.02$. 
Similarly, the crescent phase results include more contribution from Surface 1 ($0.14 \pm 0.06$) than the quadrature results, and again, the longitudinal maps agree best for slices containing majority open ocean, with the change in the maps with phase driven primarily by disagreements in the land-bearing slices.

The longitudinal mapping results from the lower S/N datasets in Fig. \ref{fig:glint_map} (lower panels) demonstrate the photometric limitations of multi-phase mapping. The middle panel with S/N=10 is quite similar to the S/N=23 case, but with slightly larger uncertainties on the crescent phase surface albedos and longitudinal maps. The lower panel with S/N=2.4 no longer captures the variability at crescent phase: the albedo and longitudinal map of both surfaces are unconstrained and significantly overlap. The convergence of the two surfaces implies that a single surface model (mean observed albedo and no map) is warranted at crescent phase. At quadrature phase, the S/N=2.4 result still shows success in extracting both surface's albedos and mapping them to the ocean and land. In the following subsection we further explore the limitations and feasibility of multi-phase longitudinal mapping, particularly with respect to difficult crescent phase observations. 

\subsection{Observational Feasibility}

We now assess the feasibility of using this phase-dependent technique to detect oceans on extrasolar planets, considering angular separation, photometric sensitivity, and target yields.  To obtain direct imaging observations of an Earth-analog exoplanet at a crescent phase of $135^{\circ}$ the planet must fall outside the coronagraph inner working angle (IWA). Additionally, the crescent phase lightcurves must be observed with sufficient photometric precision for the relatively short exposure times needed for mapping to be able to retrieve multiple surfaces. Below, we describe these observing considerations, and estimate the number of nearby planetary systems where this technique could be applied to search for oceans with upcoming telescopes. 

\subsubsection{Inner Working Angle Constraints}
\label{sec:results:iwa}

The IWA is the minimum angular separation, $\theta_{\mathrm{IWA}}$, for which a coronagraph can occult a star without blocking the light of an exoplanet in orbit about it. Since $\theta_{\mathrm{IWA}} \propto \lambda / D$, it is convenient to define 
\begin{equation}
    \theta_{\mathrm{IWA}} = N \lambda / D
\end{equation}
where $N$ is a factor set by the specific coronagraph design. Smaller values of $N$ enable planets at smaller angular separations from their host star to be imaged. In the limit that the distance to the observed system, $d$, is much farther away than the projected planet-star separation, $r$, then
\begin{equation}
    \theta_{\mathrm{IWA}} \approx r / d.
\end{equation}
We can solve for the coronagraph specific IWA parameter $N$ in terms of key telescope and system parameters, 
\begin{equation}
   N \approx 4.8 \left ( \frac{r}{1 \mathrm{AU}} \right ) \left ( \frac{D}{10 \mathrm{m}} \right ) \left ( \frac{10 \mathrm{pc}}{d} \right ) \left ( \frac{1 \mu \text{m}}{\lambda} \right ).
\end{equation}
We initially consider observations of habitable Earth-like exoplanets orbiting Sun-like stars for which $r \approx 1$ AU, at wavelengths near $\lambda \approx 0.8$ $\mu$m, but using a telescope with a currently unspecified mirror diameter, and for systems that have yet to be discovered. 

\begin{figure}
\centering
\includegraphics[width=0.47\textwidth]{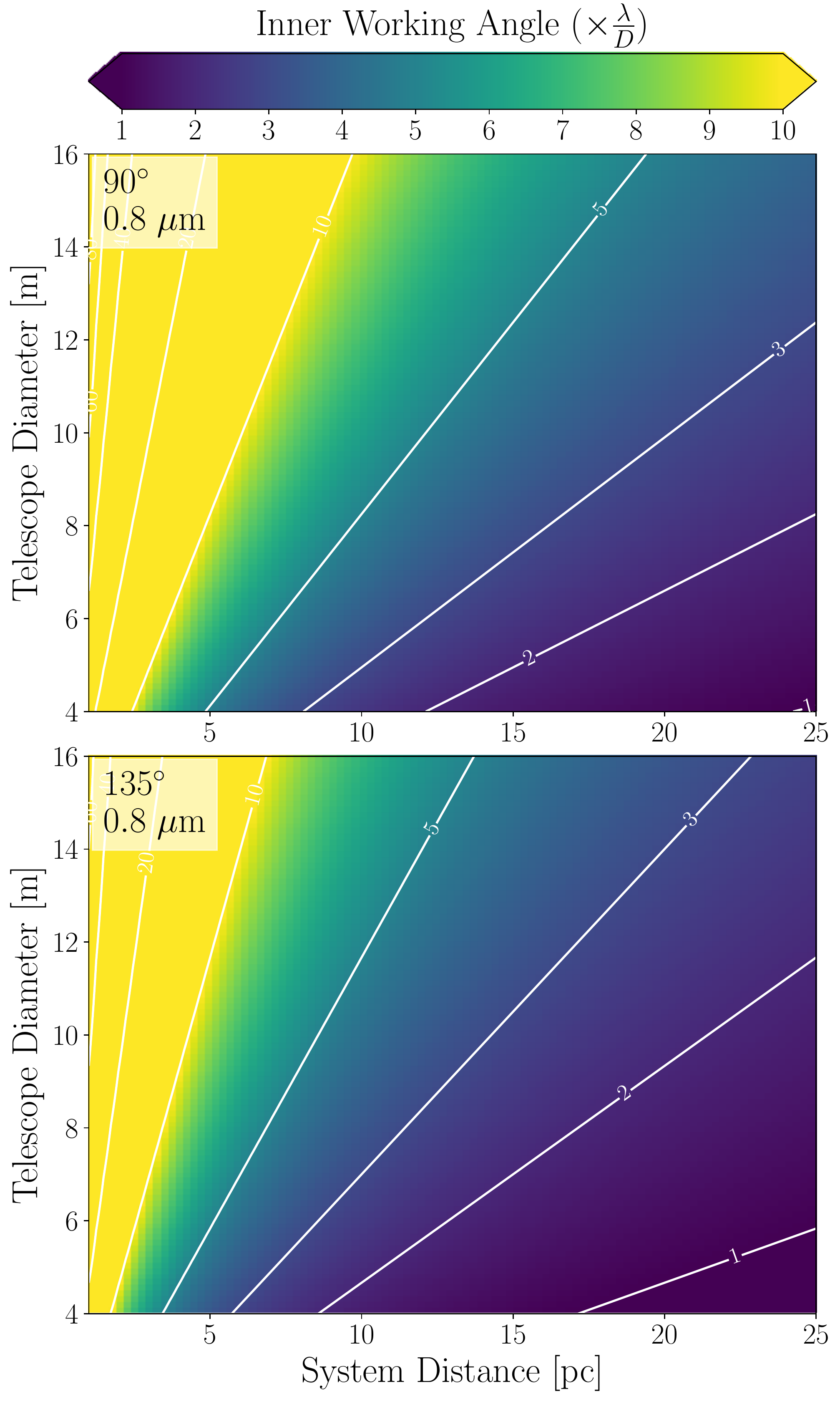}
\caption{Coronagraph inner working angle (IWA) requirements for observing Earth-analog exoplanets. The top panel corresponds to a planet at quadrature phase ($\alpha = 90^{\circ}$) and the bottom panel corresponds to a planet at crescent phase ($\alpha = 135^{\circ}$). The IWA is given as the factor which multiplies $\lambda / D$, at a fixed wavelength $\lambda = 0.8$ $\mu$m, as a function of telescope diameter $D$ and the distance to the system $d$.} 
\label{fig:iwa}
\end{figure}

To better understand the coronagraphic requirements needed for ocean detection, we explored contours of constant $N$ while varying the telescope diameter and system distance. 
Figure \ref{fig:iwa} shows coronagraph IWA requirements to directly image an Earth-analog exoplanet at quadrature phase (top panel) and at crescent phase (bottom panel). At quadrature phase the projected planet-star separation is the semi-major axis, $a = 1$ AU, but at crescent phase with $\alpha = 135^{\circ}$, the projected separation for an edge-on inclination system is $a \sin(\alpha) \approx 0.707$ AU. 
This tighter planet-star separation imposes a stricter constraint on the coronagraph IWA that ultimately reduces the distance out to which exoplanet systems may be observed at $135^{\circ}$ phase relative to quadrature phase for any given coronagraph design. For instance, a 10 meter telescope with $\theta_{\mathrm{IWA}} = 3 \lambda / D$ can image Earth analog exoplanets out to ${\sim} 20$ pc at quadrature, but only ${\sim} 14$ pc at $135^{\circ}$ degrees. 

\subsubsection{Photometric Constraints}

\begin{figure}
\centering
\includegraphics[width=0.47\textwidth]{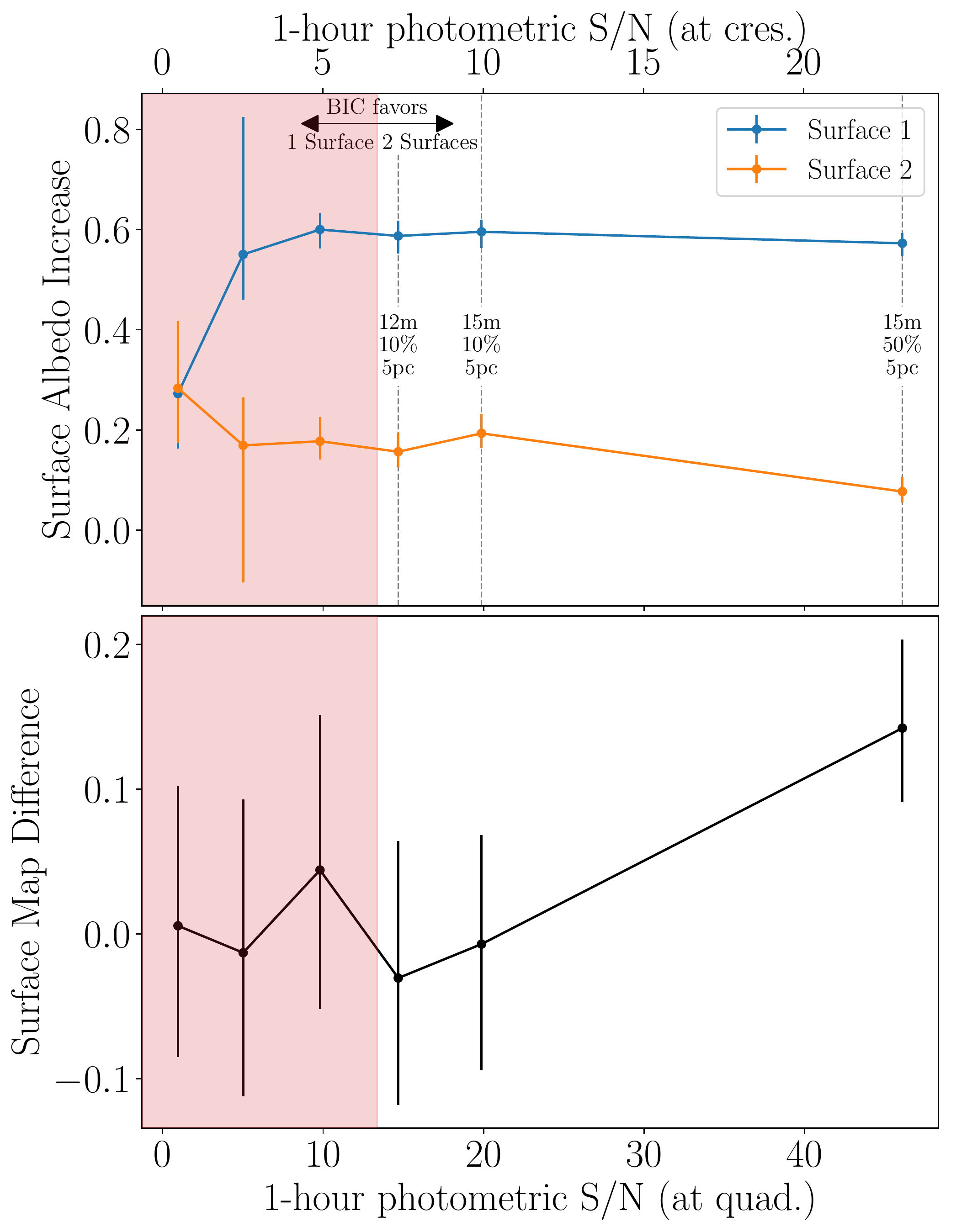}
\caption{
Change in surface albedo (top) and change in surface longitudinal area covering fractions (bottom) as a function of the signal-to-noise of the observations. The lower x-axis corresponds to the 1-hour photometric signal-to-noise for the Earth at quadrature phase, while the upper x-axis corresponds to the 1-hour signal-to-noise at crescent phase for observations of the same exoplanet, at the same distance, using the same model telescope configuration, with the same integration time. The upper panel shows the phase-dependent albedo increase of Surface 1 (blue) and Surface 2 (orange) with 1-$\sigma$ error bars derived by randomly sampling the MCMC posterior distributions. The lower panel shows a similar result but for the change in longitudinal area covering fraction. The red shaded region indicates the crescent phase photometric precision where a one surface fit (homogeneous map) is favored by the Bayesian Information Criterion (BIC) over a two surface fit.
} 
\label{fig:phase_mapping_feas1}
\end{figure}

We also considered the feasibility of detecting large changes in albedo of a particular surface with phase as a function of observational precision, and the effects of telescope design and target distance on scientific yields. We repeated the same multi-phase longitudinal mapping experiment as in \S\ref{sec:results:mapping} for the true cloudy Earth observed for 100 hours with a 1-hour cadence, but now for varying levels of photometric precision. However, instead of fixing the signal-to-noise on each lightcurve, we fixed the underlying telescope architectural parameters used to simulate the lightcurves to enforce self-consistency between observations at quadrature and crescent phase. We then determine for what data quality, or telescope design, statistically significant increases in Surface 1's albedo may be retrieved concurrent with statistically similar longitudinal maps. 

Figure \ref{fig:phase_mapping_feas1} shows both changes in retrieved apparent albedo of the two surfaces and changes in retrieved longitudinal maps as a function of data quality. The lower x-axis corresponds to the 1-hour photometric signal-to-noise for the Earth at quadrature phase, while the upper x-axis corresponds to the 1-hour signal-to-noise at crescent phase for observations of the same exoplanet, at the same distance, with the same model telescope configuration, and identical integration time. Thus, fixing all parameters except phase, crescent phase observations are approximately half the signal-to-noise of quadrature phase observations. The upper panel of Fig. \ref{fig:phase_mapping_feas1} shows the phase-dependent albedo increase of Surface 1 (blue) and Surface 2 (orange) with 1-$\sigma$ error bars derived by randomly sampling the MCMC posterior distributions. This visualization presents posterior PDFs, analogous to those shown in Fig. \ref{fig:cloudless_albedo_compare}, as points with errors along an x-axis of increasing data quality. The lower panel shows a similarly derived result, but for changes in longitudinal area covering fraction, which is analogous to results from Fig. \ref{fig:cloudless_map_compare}, but now as a function of S/N.

Figure \ref{fig:phase_mapping_feas1} indicates that the phase-dependent apparent albedo change of Surface 1 can be robustly detected and discriminated from the lack of albedo change in Surface 2 across a wide range in photometric precision, although recognizing the same map at multiple phases may prove difficult. For a quadrature (crescent) S/N $>$ 10 (5), the increase in apparent albedo with phase for Surface 1 is consistent with 0.6, with uncertainties smaller than 0.04. Over the same range, the albedo of Surface 2 slightly increases. The decrease in albedo change of Surface 2 with photometric precision suggests that greater precision aides in the recognition of Lambertian scattering surfaces. However, the opposite behavior is found for the change in maps with phase. For low S/N on the observed lightcurves the inferred longitudinal maps are consistent with no change---the maps are the same within the propagated observational uncertainties. As photometric S/N increases the maps at quadrature and crescent phase appear more unique from one another, with an average covering fraction change of $0.14 \pm 0.06$ for the highest S/N case. Similar to our findings in Fig. \ref{fig:cloudless_map_compare}, this scaling behaviour suggests that the inferred maps are actually distinct despite having the same continental distribution and sub-observer latitude. We propose that this discrepancy is due to the effect of a multi-component scattering phase function on the inferred longitudinal maps, which we discuss further in \S\ref{sec:discussion:oceans}.

The asymptotic behavior of Surface 1's apparent albedo increase for S/N $>$ 10 suggests a natural precision lower limit below which the difference in apparent albedo with phase will be lost in the noise. To further investigate this limit and determine the photometric limits of mapping---particularly at crescent phase---we find the minimum number of surfaces that are justified in fits to different S/N data. We use the Bayesian Information Criterion \citep[BIC;][]{Schwarz1978} to penalize overly complicated model fits. We take the ``optimal'' number of surfaces to be the model with the minimum BIC. We define a convention that the optimal number of surfaces determined by the BIC is the number of \textit{detected} surfaces, where a one surface fit is a homogeneous map and a two surface fit is the bare minimum to perform any meaningful mapping. This is similar to the use of principal component analysis to quantify detected surfaces \citep{Cowan2009, Fujii2017}, except we find it to be significantly more robust against observational uncertainties. 

We used the BIC to find the number of detected surfaces across a finely sampled grid in photometric S/N at both quadrature and crescent phase. For each S/N, we calculate the BIC for a one surface model by finding the best-fit from 100 randomly initialized optimizations using the \texttt{L-BFGS-B} algorithm \citep{Zhu1997}. We then increment the number of surfaces and repeat the procedure until we discover the minimum BIC. Both panels of Figure \ref{fig:phase_mapping_feas1} are shaded red to denote the crescent phase photometric precision for which only one surface is detected. The boundary occurs at a photometric precision of ${\sim} 15 \%$ (S/N=6.7) for observations at crescent phase, which we take to be the minimum precision capable of performing phase-dependent mapping on a partially cloudy Earth-like exoplanet. We also find that the boundary between one and two surfaces requires a slightly improved photometric precision of ${\sim} 10 \%$ (S/N=10) for observations at quadrature phase. This indicates that the increased color contrast between the two surfaces at crescent aids in multi-surface mapping, although higher precision measurements are easier to acquire at quadrature phase. We also note that three surfaces are detected at photometric precisions $> 2.8 \%$ (S/N=35) at quadrature. 

\begin{figure}
\centering
\includegraphics[width=0.47\textwidth]{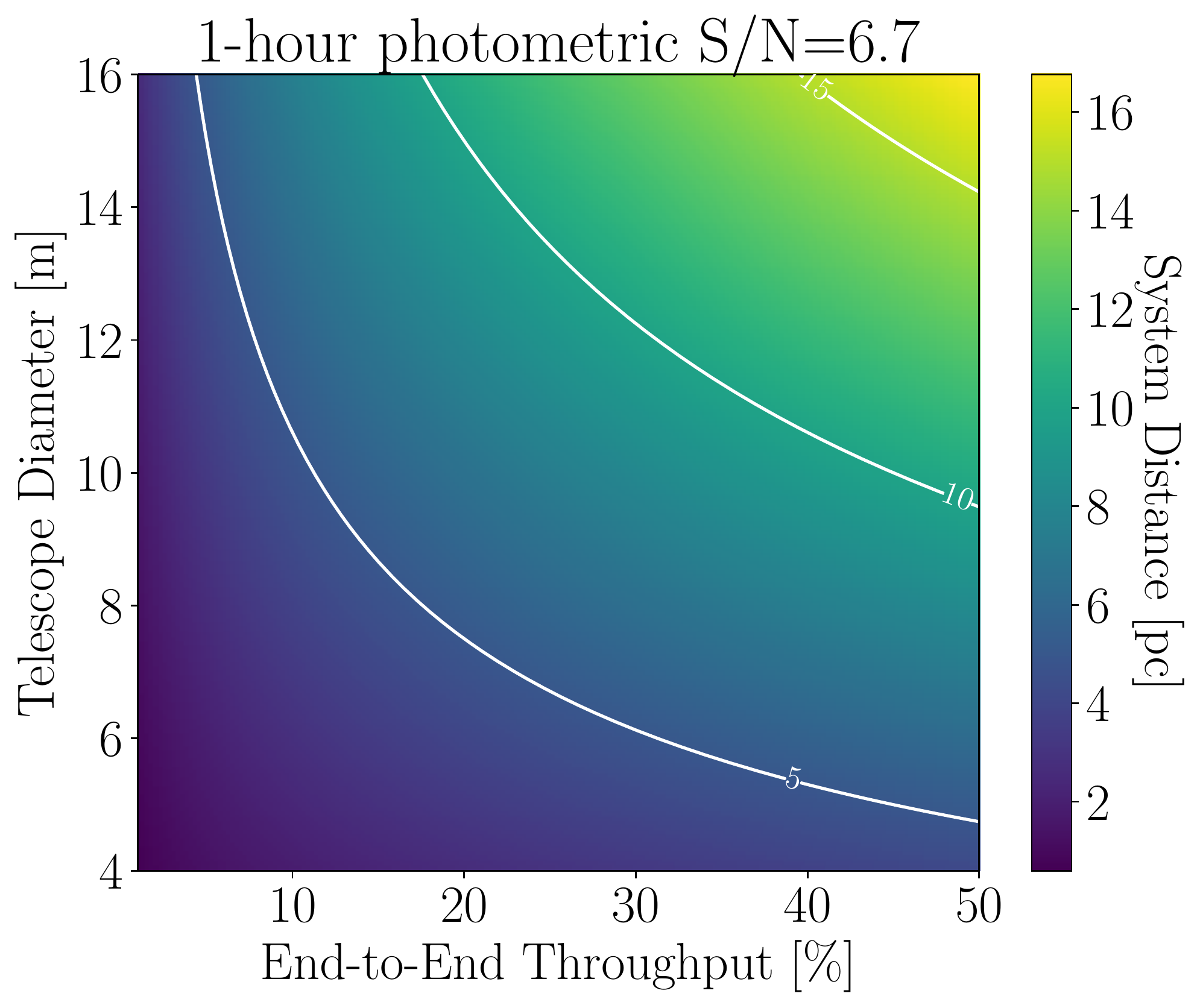}
\caption{Distance (color contours) out to which an Earth-like exoplanet could have crescent phase ($\alpha = 135^{\circ}$) observations fit with a two surface model, as a function of telescope diameter and throughput. A 1-hour photometric precision of ${\sim} 15 \%$ (S/N=6.7) is determined by the BIC to be the threshold between detecting one surface (homogeneous planet) and two surfaces (heterogeneous planet) at crescent phase, considering our observational assumptions, particularly the limited bandwidth. Minimally detectable two surface maps at quadrature phase can be performed out to approximately twice the distance shown here for crescent phase.}
\label{fig:phase_mapping_feas2}
\end{figure}

Adopting the two-surface threshold at crescent phase as the lower limit to the photometric precision, we determine the ability for different possible future telescopes to meet this requirement with respect to scientific yields. Figure \ref{fig:phase_mapping_feas2} shows the distance out to which an Earth-like exoplanet could have crescent phase ($\alpha = 135^{\circ}$) observations meet the S/N threshold for two-surface longitudinal mapping as a function of telescope diameter and end-to-end telescope throughput. For example, a 15 m with 20\% throughput could perform crescent phase mapping out to 10 pc, while an 8 m with 20\% throughput could only go out to 5.4 pc, and a 4 m out to only 2.7 pc. 

\subsubsection{Extending to Low Mass Stars} 

The forward models used here were developed for Earth-Sun analog systems. However, we can broaden our scope to consider observations of planets with surface oceans orbiting later type stars, using the Earth as a proxy for these planets.
Cooler stars have compact habitable zones and therefore smaller angular separations between planet and star compared to G dwarfs. This places a stricter constraint on the IWA to directly image exoplanets in such orbits. Using the \citet{Kopparapu2013} moist/max greenhouse HZ with the 5 Gyr \citet{Baraffe1998} stellar models, we calculate the semi-major axis of the inner and outer edge of the HZ and relate this to the IWA. 

\begin{figure*}
\centering
\includegraphics[width=\textwidth]{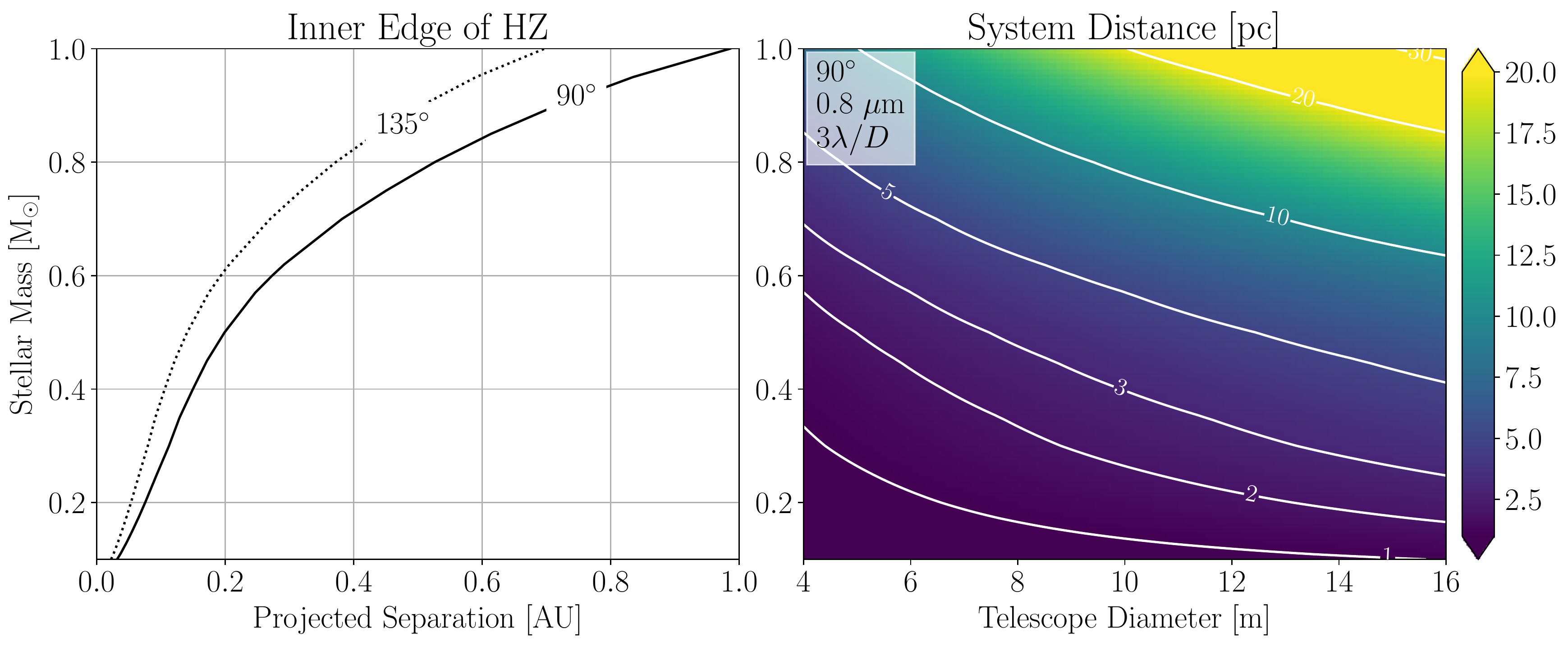}
\caption{\textit{Left:} Projected separation between star and planet at the inner edge of the habitable zone (moist greenhouse limit) for low mass stars. Quadrature and crescent ($135^{\circ}$) phase projected separations are shown as solid and dotted lines.  \textit{Right:} Maximum distance (color contours) out to which a planet at the inner edge of the HZ can be directly imaged as a function of stellar mass and telescope diameter. Phase, wavelength, and IWA are fixed at $90^{\circ}$, 0.8 $\mu$m, and $3 \lambda / D$. Distances can be readily scaled to the apparent phase $\alpha = 135^{\circ}$ by multiplying by $\sin(\alpha)$. }
\label{fig:other_stars_iwa}
\end{figure*}

Figure \ref{fig:other_stars_iwa} shows the distance out to which a planet at the inner edge of the HZ (and therefore the entire HZ) may be directly imaged as a function of telescope diameter and host star mass, for $90^{\circ}$ phase, 0.8 $\mu$m, and IWA of $3 \lambda / D$. A 15 m telescope can directly image planets at the inner edge of the HZ of a 0.6 M$_{\odot}$ K dwarf at $90^{\circ}$ and $135^{\circ}$ out to 8.3 pc and 5.8 pc, respectively. A 9-meter coronagraph observing the same system would only be able to go out to 5.0 pc and 3.5 pc for phases of $90^{\circ}$ and $135^{\circ}$, respectively; a 6-meter out to 3.3 pc and 2.2 pc, respectively; and a 4-meter out to 2.2 pc and 1.6 pc, respectively. Note that the stellar distances for accessing the inner edge of the HZ given in Fig. \ref{fig:other_stars_iwa} are conservative compared with outer edge calculations, which we also account for next.  

\subsubsection{Yield Estimates}

\begin{figure*}
\centering
\includegraphics[width=\textwidth]{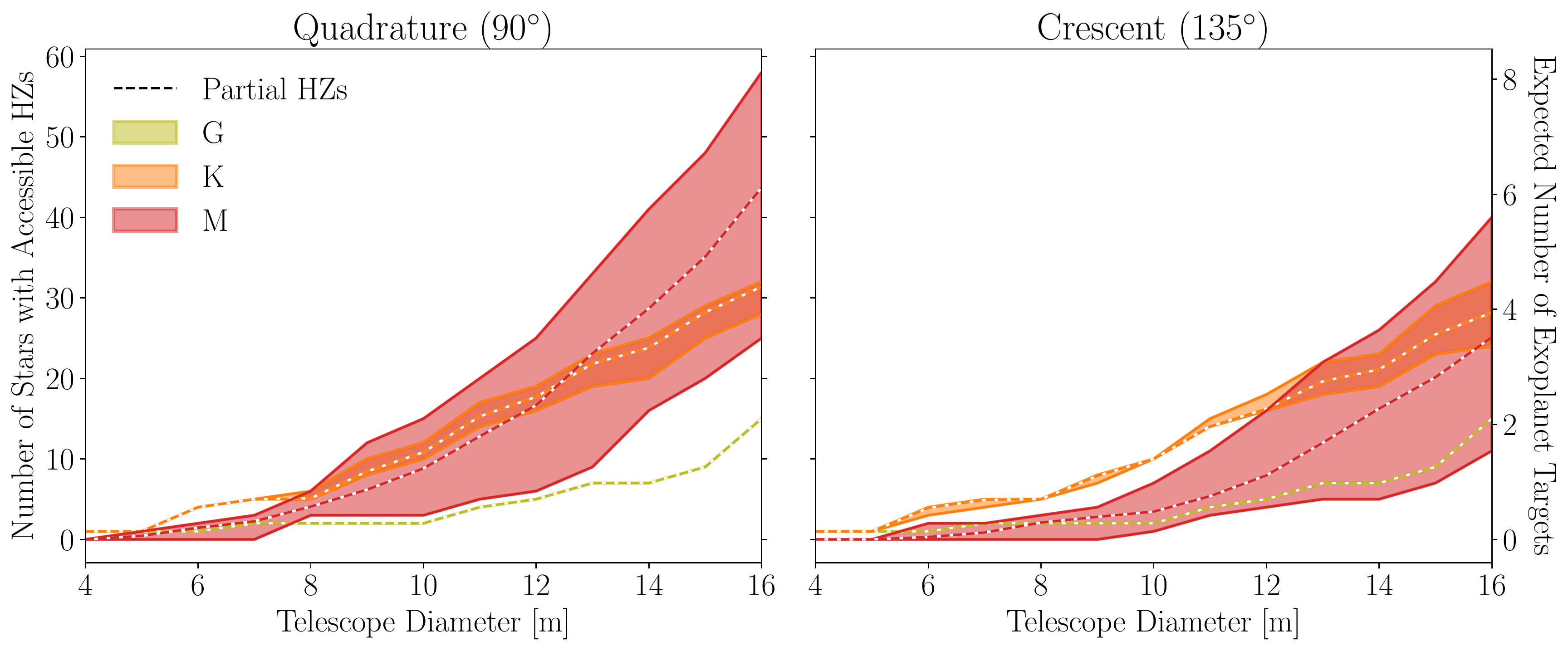}
\caption{
Yield estimates for observations of HZ exoplanets at quadrature (left panel) and crescent phase (right panel) as a function of telescope diameter for nearby G, K, and M dwarfs. Systems are counted if a hypothetical planet in the HZ falls outside of the IWA (and is therefore observable), assuming $\theta_{IWA} = 3 \lambda / D$, and if the 1-hour S/N at crescent phase exceeds our photometric threshold of 6.7 (see Fig. \ref{fig:phase_mapping_feas1}), assuming a $20\%$ throughput. Both panels share the same y-axes, where the left y-axis gives the number of stars that would have observable HZ planets. The right y-axis gives the expected number of exoplanets after considering the geometric probability of passing through $135^{\circ}$ phase ($70\%$) and assuming a $20\%$ occurrence rate for planets in the HZ. The shaded regions are bounded below and above by the inner and outer edges of the HZ, respectively, and demonstrate a sensitivity in yield depending on whether the observed planet is at the inner or outer edge of the HZ. The dashed lines correspond to the sum of all partially observable HZs (e.g. half a count for a star with half of the HZ observable outside the IWA). 
Observations of HZ exoplanets are limited by photometric precision for G dwarfs while M dwarfs are significantly more limited by the IWA. 
}
\label{fig:stellar_yield}
\end{figure*}

We now provide initial estimates for how many planets might be accessible to ocean detection, as a function of telescope diameter.
We use the Nearby Stellar Systems Catalog for Exoplanet Imaging Missions \citep[ExoCat;][]{Turnbull2015} to estimate target yields for G, K, and M dwarfs in the stellar neighborhood. For each star in ExoCat we use the stellar effective temperature and luminosity to calculate the moist greenhouse inner edge of the HZ and maximum greenhouse outer edge using relations from \citet{Kopparapu2013}. We then relate the projected separation of hypothetical exoplanets at the inner and outer edges of the HZ to the IWA of telescopes of different diameters, assuming $\theta_{IWA} = 3 \lambda / D$, to determine if a planet in such an orbit is observable. 
Additionally, to meet the photometric requirements for mapping, we also check to see if an Earth-size and albedo planet at the inner edge of the HZ exceeds our 1-hour photometric S/N threshold of 6.7, assuming the same telescope and instrumental specifications provided above and a $20\%$ throughput. If a given stellar system satisfies both the IWA and the photometric conditions then we count that system as observable. 

However, not all stars will have a HZ planet, and not all HZ planets will be optimally inclined so that they pass through a crescent phase of $135^{\circ}$. These are questions on the occurrence rate of HZ exoplanets and the random distribution of system inclinations, both of which are fundamental to exoplanetary target yield estimates. Numerous studies have attempted to infer the occurrence rate of Earth-like planets in the HZ of different stellar types \citep[e.g.][]{Dressing2013, Kopparapu2013b, Foreman-Mackey2014, Burke2015}, but large uncertainties persist for temperate rocky worlds. We assume an occurrence rate of $20\%$ for G, K, and M dwarfs, which is consistent with published values. Although trends in occurrence rate with stellar type may exist \citep{Mulders2015}, we elect to keep our calculations simple so that they may be easily scaled as occurrence rates are refined. The prior probability of randomly oriented system inclinations is $\sin(i) di$ \citep{Luger2017}. This means that $70\%$ of randomly oriented systems will have planets tracing orbital paths that pass through a phase of $135^{\circ}$ ($i > 45^{\circ}$). We use the $20\%$ occurrence rate and the $70\%$ random inclination prior to scale our potential stellar targets into an expected number of exoplanets for which oceans may be searched for. 

The left y-axis of Fig. \ref{fig:stellar_yield} shows the number of stars for which an assumed optimally-aligned planet ($i = 45^{\circ}$ - $90^{\circ}$) in the HZ of G (yellow), K (orange), and M (red) dwarfs could be observed at phase angles of $90^{\circ}$ (left panel) and $135^{\circ}$ (right panel) as a function of telescope diameter. The right y-axis of Fig. \ref{fig:stellar_yield} shows the expected number of HZ exoplanets that could potentially be searched of oceans using our crescent phase glint mapping technique. These observations are assumed to be at $0.8$ $\mu$m, using a coronagraph with an IWA of $3 \lambda / D$. 
The shaded regions in Fig. \ref{fig:stellar_yield} are bounded below by the inner edge of the HZ and above by the outer edge. The dashed lines correspond to the sum of all partially observable HZs, where a partial HZ is calculated as the radial fraction of the HZ that is observable beyond the IWA. Thus the dashed lines fall between the corresponding shaded region because 100\% of the HZ is observable if the inner edge falls outside the IWA, while $0\%$ of the HZ is observable if the outer edge is equal to the IWA.

Figure \ref{fig:stellar_yield} suggests that K and M dwarfs will make up the bulk of the stellar hosts for surface ocean assessments. Low mass K and particularly M dwarfs make up the bulk of the nearby stellar population. However, these stars have compact HZs that are difficult to observe in direct imaging. Additionally, the planet-to-star contrast ratio increases towards lower mass, fainter stars. This decreases the noise due to unocculted stellar light (speckles), particularly for systems with contrast ratios that are lower than the design contrast of the direct imaging mission ($10^{-10}$ assumed here). Fig. \ref{fig:stellar_yield} suggests that K dwarfs strike a balance between these effects for the smaller telescope diameters considered, and may be optimal stellar hosts for direct imaging studies of HZ planets. G dwarfs have appreciably separated HZs, so that if the systems are near enough to achieve a sufficient photometric precision, then the IWA will not be a restricting factor for crescent phase observations. This is in agreement with our earlier findings. M dwarfs occupy a different domain: their favorable planet-to-star contrast ratios more easily satisfy S/N requirements, so that IWA is instead more limiting on both the stellar distance out to which M dwarfs may have their HZs observed, and crescent/gibbous phase observations of any HZ planet.

Considering the IWA and S/N constraints, we find that 1 to nearly 10 HZ planets could be assessed for signs of time-varying ocean glint signals with telescopes in the 6 to 15 m diameter range.
The nearby single stars Tau Ceti, Epsilon Eridani, and Lacaille 8760 (AX Microscopii) are the respective G, K, and M dwarfs that have the largest angular separation to the inner edge of the HZ, making them quite optimistic targets, considering IWA constraints. 
Telescopes smaller than 12 m may have the potential for crescent phase observations around more K dwarfs than either G or M dwarfs. 
Telescopes larger than 12 m may have more M dwarfs targets than K dwarf targets. 
A 15 m with an IWA of $3 \lambda / D$ could perform crescent phase $0.8$ $\mu$m direct imaging in the HZ (including the sum of partially observable HZs) of approximately 9 G, 25 K, and 20 M stars. A 12 m could do the same for 5 G, 16 K, and 8 M stars; a 9 m for 2 G, 8 K, and 3 M stars; a 6 m for 1 G, 4 K, and 0 M stars; and a 4 m for 1 G, 1 K, and 0 M stars.
Considering random system geometries and assuming an occurrence rate, as discussed above, we find that a 15 m, 12 m, 9 m, 6 m, and 4 m will have approximately 8, 4, 2, 1, and 0 expected planet(s), respectively, in the HZ amenable to crescent phase mapping investigations.

\section{Discussion} \label{sec:discussion}

We investigated the unique time- and phase-dependent observables of Earth oceans in the context of direct habitability assessments for terrestrial exoplanets. In the following subsections we discuss how our findings may be used to increase the robustness of ocean detection on exoplanets (\S\ref{sec:discussion:oceans}) and assess the potential yield and required observational conditions for ocean searches on extrasolar planets (\S\ref{sec:discussion:observe}).    

\subsection{Increasing the Robustness of Exoplanet Ocean Detection}
\label{sec:discussion:oceans}

We find that the robustness of exo-ocean detection can be increased via solutions to the apparent albedo and longitudinal map of multiple surfaces observed at multiple orbital phase angles.  We retrieved the longitudinal map and apparent albedo of two unique surfaces from simulated lightcurves of Earth at quadrature, crescent, and gibbous phase. At quadrature and gibbous phase, one low albedo surface is mapped to longitudinal slices that we know from Earth to be ocean bearing, while the other more reflective surface is mapped to the continents. At crescent phase, the maps of the dominant surfaces appeared quite similar to the quadrature phase solutions, and the albedo of the land-containing slices was nearly the same as at quadrature, but the apparent albedo of the ocean-dominated slices exhibited a large and discernible increase in albedo of ${\sim} 0.5$. 
This sizable increase in apparent albedo of the ocean-bearing longitudinal slices seen at crescent phase comes from the ocean glint spot. Since our inverse model assumes surfaces are Lambertian scatterers, deviations from Lambertian are encoded in the inferred apparent albedo of the surface, or the albedo that the surface must have (when integrated over the longitudinal slice) to reproduce the observed flux. The fact that the longitudinal maps of each surface appear quite similar at quadrature and crescent phase provides a means to identify that the same surfaces are mapped. This helps to attribute changes in surface reflectivity to specific surface components. The planetary surface that had appeared dark at quadrature, must reflect light more efficiently at crescent phase. Such a defined deviation from Lambertian behavior is strongly suggestive of ocean glint. 

Although our cloudless Earth tests demonstrated that glint effects can be extracted from crescent phase longitudinal mapping, our tests with realistic time-varying clouds revealed limitations and caveats for the technique. Clouds made it much more difficult to extract the same map at both quadrature and crescent phases. Crescent phase observations are more sensitive to clouds than more nadir viewing geometries. Clouds therefore inhibit, but do not fully prevent, accurate surface longitudinal mapping at crescent phase. We were still able to detect the presence of large landmasses, like Africa, which effectively interrupted the glint spot. 

However, clouds had minimal impact on the surface albedo results with phase. Despite the added difficulty for accurately detecting the same map, we did confidently detect the increase in albedo of the ocean-bearing slices, along with the lack of albedo increase of the land-bearing slices. Both retrieved surfaces exhibited a slight increase in albedo going from quadrature to crescent phase when clouds were included. Without clouds Surface 2 (land) did not increase in inferred albedo ($0.02 \pm 0.02$) with phase, while with clouds it increased by about 0.1 ($0.08 \pm 0.02$), a $30 \%$ increase. Surface 1 (ocean) without clouds increased in albedo by 0.5 ($0.47 \pm 0.02$) with phase, a factor of ${\sim}20$ increase, while with clouds it increased by 0.6 ($0.57 \pm 0.02$), a factor of ${\sim} 5$ increase with phase. This added increase of 0.1 in the apparent albedo of both surfaces towards crescent is due to forward scattering by clouds, which we find imparts a significantly smaller phase-dependent effect than ocean glint \citep[c.f.][]{Robinson2010}, and the effect of glint strongly correlates with only one of the surface components. \citet{Cowan2009} also noted that clouds appeared to increase the albedo of both principle components used in their study, rather than appearing as a unique principle component (or surface). 
As a result of decomposing the lightcurves into two surfaces, we see a much stronger effect from glint (${\sim} 5 \times$) than that noted by \citet{Robinson2010} (${\sim} 2 \times$), which could make this technique a more sensitive tool for ocean detection than attempting to observe phase-dependent brightness in the disk-average.  

Longitudinal mapping at crescent phase naturally breaks the latitude-albedo false positive for ocean glint \citet{Cowan2012} by attributing the global albedo increase observed at crescent phases to a specific surface on the planet. High S/N, high cadence, reflected light observations at crescent and quadrature (or gibbous) phase enable the longitudinal map and albedo to be retrieved. The time-variability of a rotating planet's disk-integrated light is used to deconvolve the relative albedo increase seen towards crescent phase into specific surfaces that are responsible for the relative increase. This may increase the confidence that high latitude ice is not responsible for the relative albedo increase. Furthermore, the boosted strength of the glint signal found after the deconvolution (${\sim}5 \times$) exceeds the likely increase in disk-integrated albedo that is expected from the latitude-albedo effect for an Earth-like planet with polar ice caps \citep[${\sim} 2 \times$; ][]{Cowan2012}.  

We find that phase-dependent deviations in the longitudinal maps of, our ideal case, the cloudless Earth further implicate non-Lambertian surface scattering behavior. The difference between the quadrature and crescent phase maps in Fig. \ref{fig:cloudless_maps} can be attributed to the footprint on the planet that contributes the most flux at each phase, and therefore drives our inferences. \citet{Cowan2013b} refer to this as the kernel of convolution, which has a longitudinal and latitudinal extent determined by the system geometry and phase function of the reflecting surface \citep{Schwartz2016}. At quadrature phase, the planet's surface is well described by a Lambertian surface with a kernel that extends well above and below the equator (for our case here with a 90$^{\circ}$ inclination and 0$^{\circ}$ obliquity). However, at crescent phase the ocean deviates from diffuse Lambertian scattering as light reflected off the ocean is dominated by the glint spot. The effective size of the glint spot is significantly more concentrated in both longitude and latitude compared with the Lambertian kernel of convolution. Consequently, the ocean glint probes a relatively small latitudinal extent. Thus, inferred glint maps are highly sensitive to interruptions (e.g. land, ice, and clouds) along the direct path that the glint spot takes across the surface of the planet. In our test case of Earth with 90$^{\circ}$ inclination and 0$^{\circ}$ obliquity, the dominant latitude and path of the glint spot is the equator. 

The effect of kernel size helps to explain why our cloud-free longitudinal maps at quadrature phase (Fig. \ref{fig:cloudless_maps}) show a sensitivity to continents in the mid-latitudes, such as North America and Asia, which does not appear in the crescent phase maps. These continents never intersect the path of the latitudinally compact glint spot, and are therefore not mapped at crescent phase. Thus, the latitudinal heterogeneity of the planet leads the inferred maps to slightly disagree between quadrature and crescent phase. This does not occur strictly due to the system geometry as explored in \citet{Schwartz2016}, but instead because the scattering phase function of the planetary surface(s) affects the width of the convolution kernel. Conversely, the differences found between the quadrature and gibbous phase maps are driven more by the illumination geometry, which favors smoother longitudinal maps at gibbous phase because more illuminated longitudes are integrated together, thus decreasing spatial resolution in the maps \citep{Fujii2017}.  


For demonstrative simplicity we have considered planets with zero obliquity and in an edge-on inclined orbit. Since the sub-observer latitude does not change with phase, this is the optimal case for detecting the same longitudinal map at multiple phases. However, even for this optimistic case with the same latitude dominating the peak reflected flux, we had difficulty recognizing the same map at multiple phases due to latitudinal heterogeneity, as discussed above. 

This presents the possibility of false negatives that may arise due to the difficulty of inferring the same longitudinal map at multiple phases. If the observed planet has a prominent north-south continental dichotomy and a non-zero obliquity, like Earth, or a non-edge-on inclination, then time-series observations at multiple phases will constrain the two-dimensional map of the planet \citep[e.g.][]{Kawahara2010, Kawahara2011, Fujii2012, Berdyugina2017} and the obliquity \citep{Schwartz2016, Farr2018}. For these cases it will be more difficult to distinguish between changes with phase and changes in surface type. As a result, multi-phase longitudinal mapping will perform best for system geometries that are not amenable to two-dimensional spin-orbit mapping. However, exoplanets with non-zero obliquity may still exhibit strong lightcurve variations at crescent phase that are best explained by the combination of a Lambertian and a non-Lambertian component. It may be that congruence of the multi-phase maps is not necessary to place strong constraints on the detection of glint. Indeed, our finding that the glint signal is found in only one inferred surface, and that the glint effect far exceeds forward scattering by clouds, may favor an ocean glint interpretation, albeit with less confidence than if the multi-phase maps matched. 

Even the detection of glint is not uniquely indicative of surface liquid water. Surface liquid reservoirs of any composition may cause a specular reflection due to their mirror-like smoothness. This is perhaps best exemplified by the glint detection from Titan's hydrocarbon lakes \citep{Stephan2010}. Freshly-formed smooth ice may also deviate from Lambertian scattering \citep{Williams2008}. Furthermore, terrestrial glint has been identified from light scattering off oriented ice crystals in high altitude cirrus clouds \citep{Kolokolova2010, Marshak2017}. Such signals could be distinguished from surface glint due to their lack of reproducibility on subsequent rotations, or by their anti-correlation with water vapor absorption features in the spectrum that would indicate a cloud-truncated path length. 
These cases exemplify why additional context is necessary for conclusions to be drawn on the surface liquid composition, such as surface pressure and temperature constraints \citep{Robinson2018}, or the phase-dependent relative change in the surface albedo as discussed in this work.

Because the VPL Earth Model only accurately reproduces Earth, our study only considered Earth's true geography, rotation rate and atmospheric composition. Future studies should generalize the technique of glint mapping with randomly sampled rotation rates and geographies. Furthermore, we assumed that the rotation rate was precisely known. However, a simple periodogram analysis indicates that our time-series data used for longitudinal mapping with a one-hour cadence and 100-hour duration is more than sufficient to confidently infer the rotation rate of Earth. In general, when the S/N is sufficient for longitudinal mapping to favor two or more surfaces and the observational duration exceeds the rotation rate, then the rotation rate can reliably be retrieved. This agrees with the findings of \citet{Palle2008} and \citet{Fujii2014}.  However, the rigorous treatment of directly imaged terrestrial exoplanet rotation rates is beyond the scope of this paper, and we leave it for follow-up work. 

Additionally, closer-in planets are more likely to be in tidally locked rotation states, for which our assumption that the rotational period is much shorter than the orbital period breaks down. Further work should investigate the potential for glint detection in reflected light phase curve measurements of tidally locked exoplanets in the habitable zones of late K and M dwarfs. In these cases the interruption of a glinting ocean by land may manifest as a strong asymmetry in the phase curve between the waxing and waning crescent phases.

Using the apparent albedo assumes that the illumination phase angle and planet radius are known, but in practice, both of these quantities will be difficult to precisely constrain for directly imaged exoplanets.  This is due to uncertainties in orbit determinations and the radius-albedo-phase degeneracy \citep{Nayak2017}. 
Planet size cannot be measured unless the planet is transiting, and it is highly unlikely that the few planets amenable to glint mapping observations have transiting orbits.  
Alternatively, constraints on the planet's radius may be derived for non-transiting planets from a combination of photometric measurements and constraints on the planet's albedo, radius, and atmospheric composition determined via spectrum retrieval \citep[e.g.][]{Nayak2017, Feng2018}.
Multi-epoch observations can be used to derive the planet's orbit, either prior to, or using the multi-epoch observations studied in this paper. The radius-albedo degeneracy could potentially be broken by relating the inferred surface composition to it's laboratory measured reflectance spectrum, particularly if phase-dependent glint were to provide a strong indication on the composition of the reflecting surface. Similarly, identifying the slight increase in reflectivity towards crescent for both surfaces could indicate forward scattering by clouds and may be used to rule out a low planet albedo, thereby helping to constrain the planet radius. 
In the event that the planet's radius is unconstrained or incorrectly assumed, multi-phase observations may still reveal \textit{relative} increases in the apparent albedo of multiple surfaces with the caveat that the apparent albedo and longitudinal area covering fraction of each surface are not precisely known. 

A number of improvements to the fitting procedures used in this work could be made in the future.
We fit the light curves at each phase independently, but multiple epochs can instead be fit simultaneously \citep[e.g.][]{Farr2018}, while allowing for the surface albedo to change with phase. 
A stronger prior could be enforced that only one of the surfaces is allowed to change albedo as a function of phase. This would turn the physical prior that land does not glint into a model constraint, thereby reducing the number of fitting parameters and helping with lower S/N cases.
Additionally, a Gaussian process could be used to enforce smoothness in the longitudinal maps and allow for the characteristic angular length scale to be inferred \citep[e.g.][]{Farr2018}. This could provide enough flexibility in the mapping procedure to more accurately assess the congruence of maps at multiple phases.  


Our work has extended rotational mapping to multiple observed planetary phases to combine specific observables of exoplanetary habitability. Previous studies have demonstrated the utility of multi-wavelength lightcurve inversions to map directly imaged exoplanets in one-dimension \citep{Cowan2009, Fujii2010, Fujii2011, Cowan2013a}, and two-dimensions \citep{Kawahara2010, Fujii2012, Berdyugina2017, Farr2018}. While previous two-dimensional exoplanet mapping concepts have leveraged the planet's obliquity and/or inclination to scan latitudes using observations at multiple orbital epochs \citep{Schwartz2016}, we have instead considered phase-dependent effects for observations at multiple epochs for cases where negligible obliquity and a nearly edge-on planet orbit would prevent two-dimensional mapping. Our crescent phase simulations of Earth demonstrate a clear increase in reflectively relative to more illuminated phases due to specular reflection off the oceans, as seen in previous studies \citep{Williams2008, Robinson2010, Robinson2014}. The variability of Earth increases towards crescent phase as the illuminated crescent may be more completely dominated by a single end-member surface spectrum \citep{Oakley2009, Robinson2010}. We demonstrated for the first time that crescent phase variability can be leveraged to map ocean glint. 

\subsection{Observational Feasibility and Yield Estimates}
\label{sec:discussion:observe}

The observations discussed above will require a telescope that can achieve simultaneous multi-wavelength, time-series, direct imaging observations of an Earth-like exoplanet at crescent phase. This will require a next-generation, high contrast, space-based direct imaging mission equipped with a coronagraph or a starshade. For glint-mapping observations to be feasible, light from the planet must fall outside the coronagraph or starshade inner working angle and at sufficient signal-to-noise to justify spectral unmixing inverse methods, without overfitting. 

We find that the IWA constraints that must be overcome to directly image an Earth-analog at a crescent phase of $135^{\circ}$ are less stringent than the constraints on the photometric precision. A 9 m, 12 m, and 15 m space-based coronagraph with an IWA $< 3 \lambda / D$ would be capable of observing an Earth-Sun analog at a $135^{\circ}$ phase angle out to ${\sim}$13, 17, and 21 pc, respectively. However, a 9 m, 12 m, and 15 m telescope with a 20\% end-to-end throughput can only reach the requisite ${\sim} 15\%$ 1-hour photometric precision on multi-wavelength lightcurves of Earth at $135^{\circ}$ phase out to ${\sim}$6, 8, and 10 pc, respectively. 

We estimate that there will be potentially habitable exoplanets orbiting nearby stars that will be amenable to glint searches using a next-generation, space-based telescope. The IWA, which is less problematic than photometric precision for G dwarfs, becomes a more serious concern for later type stars where the HZ is more compact. On the other hand, the larger population of nearby K and M dwarfs compared to G dwarfs, and the more favorable planet-star contrast ratios for these later type stars both help to counteract the IWA concern.  This allows for the possibility of quadrature and even crescent phase observations of G, K and M dwarf HZ planets using coronagraph-equipped telescopes with diameters in the 6-15 m range. Telescopes with diameters approximately ${\ge}8$ m may have more K and M dwarf targets than G dwarf, depending on how exoplanet occurrence rates scale with stellar type. 

Since performing longitudinal mapping at multiple phases requires relatively high S/N observations, then larger mirror size, and/or improvements in optical throughput and coronagraph spectral bandwidth could increase the potential number of oceans detected. Our derived 1-hour S/N requirement assumed 100 hours of consecutive observations at each phase, however, lower S/N with more exposures or higher S/N with fewer exposures could still achieve qualitatively similar results. This could potentially open the door to glint mapping observations for smaller telescopes, such as a 4 m, but would require more than 100 hours of observation at each phase, and can likely only search for oceans on planets orbiting in the HZ of Alpha Centauri A and/or B, due to their proximity.  

Although our study specifically modeled coronagraph-equipped telescopes, our results are
also applicable to external occultors, such as a starshade. The main difference between coronagraph modeling and starshade modeling for the purposes of this study are the IWA, throughput, and spectral bandwidth. 
Although starshades can attain $\lambda/D$ IWA, they are currently being considered for smaller telescopes \citep[${\le} 4$ m;][]{Seager2015, Spergel2015, Mennesson2016} than those using only an internal coronagraph. 
Therefore, a 4m-class starshade-enabled telescope with an IWA of $\lambda/D$ would be comparable to an 8m with internal coronagraph and an IWA of $2 \lambda/D$. 
For equivalent IWAs the higher yield of possible ocean observations will go to larger aperture and/or enhanced throughput, and external occultors more readily achieve higher throughput than internal occultors due to fewer optical elements \citep{Mennesson2016}.
Additionally, starshade technology permits larger spectral bandpasses to be simultaneously observed. This will allow more photons to be collected, effectively increasing the throughput, and will also provide a broader wavelength range over which to identify unique planetary surface albedo components.
Figures \ref{fig:iwa} and \ref{fig:phase_mapping_feas2} can easily be interpreted for a telescope with a starshade by considering IWAs, diameters, and throughputs that are more applicable to such an observatory. 
However, the multi-epoch approach discussed in this work will require two pointings at the same target, possibly months apart, which is more constrained by the formation flying operations of the starshade.

Due to contrast considerations, large aperture ground-based telescopes will be unlikely to observe habitable planets around G dwarfs in reflected light, but may excel for K and M dwarfs. 
Ground-based 30 meter class telescopes have an estimated coronagraph raw contrast of $10^{-7} - 10^{-8}$ \citep{Guyon2005, Pueyo2013}, compared with estimates for space-based observatories of ${\sim} 10^{-10}$. 
Since Earth-Sun analogs have a flux contrast in the visible of ${\sim} 10^{-10}$, ground-based observations will face $10^2 - 10^3$ times more noise from an inability to fully block stellar light, which will more than erase gains from larger diameter mirrors. However, ground-based direct imaging of nearby habitable zone planets orbiting K and M dwarfs may be optimal for such 30-m class facilities due to more favorable planet-star contrast ratios, and the larger aperture helps to ameliorate some IWA concerns for these compact systems. 

While similar previous studies have focused on broad wavelength coverage \citep[e.g.][]{Cowan2009}, we focused on a narrow wavelength region (0.68 - 0.8 $\mu$m) to remain relevant to current coronagraph design that is unable to effectively control the wavefront over broad wavelength ranges \citep{Pueyo2013}. As a result of the narrow spectral coverage and the intrinsic faintness of target habitable exoplanets, we simulated multi-wavelength, time-series photometry by rebinning hourly spectroscopy. This allows for simultaneous multi-wavelength, time-series without the need to change photometric filters. This also means that the entire duration of the time-series may be coadded into one spectrum that contains the disk- and time-averaged spectrum of the planet, which would be an excellent data product for atmospheric retrieval studies \citep[e.g.][]{Feng2018}. Thus, we argue that time-series spectroscopy is more favorable than traditional filter photometry for the time intensive observations discussed in this paper. 
Low noise detector technology will enable this type of time-dependent spectroscopic observation. Microwave Kinetic Inductance Detectors (MKIDs) are a promising new technology that directly measure photon energy as they excite the detector, allowing for simultaneous photon counting and low-resolution spectroscopy \citep{Meeker2018}. Such advances would naturally allow for simultaneous multi-wavelength time-series observations. 
Furthermore, the wavelength range that we considered (0.68 - 0.8 $\mu$m) contains the 0.76 $\mu$m O$_2$ A-band and the vegetation red-edge---two potential biosignatures---and thus our IWA results are also directly applicable to observing biosignatures on potentially habitable nearby exoplanets. Ultimately, the atmospheric context necessary to more confidently characterize the habitability of an exoplanet target may be contained within the same dataset used for the mapping results presented in this paper, but along the spectral-, rather than temporal-dimension.

\section{Conclusions} \label{sec:conclusion}

We successfully mapped Earth as an exoplanet using simulated multi-wavelength, time-series observations at multiple orbital phases, while including ocean glint and forward scattering by clouds. We confirmed the findings of \citet{Cowan2009} that planet mapping in the specular reflection regime can identify two surface components corresponding to oceans and continents. However, our results extend this work to show that the specularly reflecting glint spot dominates the inferred crescent phase maps, just as a dark blue surface dominates our quadrature maps for ocean-dominated worlds. Interruptions in the crescent phase glint signal as Lambertian scattering continents rotate into view leads to a strong blinking effect, which allows for separate longitudinal maps to be constructed of Lambertian and non-Lambertian surfaces. We identified that the partitioning of Lambertian and non-Lambertian surfaces between retrieved surfaces is in stark contrast to the effects of forward scattering by clouds, which introduces a global apparent albedo increase for all surfaces towards crescent phase. Such multi-epoch studies may be used to detect the strong deviations from Lambertian scattering that oceans exhibit and thereby increase the robustness of ocean detection on directly imaged exoplanets. 

We found that recognizing the same longitudinal map at multiple phases will be difficult for a variety of reasons. Clouds introduce uncertainty into the surface maps; non-zero obliquities and inclinations ${<} 90^{\circ}$ cause latitudinal sensitivities with phase; and the effective latitudinal extent of the mapped longitudes is a function of phase. However, even cases with distinct longitudinal maps at each phase may exhibit a time-dependent rotationally-induced glint signal, which differs significantly from forward scattering by clouds. 

We estimate that space-based, high-contrast telescopes with diameters larger than 6-m will have at least one potentially habitable target amenable to glint mapping via crescent phase observations, with a 15-m telescope offering nearly 10 targets. Although the technique of glint mapping discussed here is not applicable to all potentially habitable planets, it may be a key tool for increasing the robustness of habitability assessment for the most promising direct imaging targets. The discovery of an ocean on an exoplanet will transform exoplanet science by unveiling a truly habitable planet beyond Earth, bringing us one step closer to answering fundamental questions on the prevalence of life in the universe.

\acknowledgements

We gratefully acknowledge useful discussions with C. Stark, G. Arney, and D. Fleming. This work was supported by the NASA Astrobiology Institute's Virtual Planetary Laboratory under Cooperative Agreement number NNA13AA93A. This work made use of the advanced computational, storage, and networking infrastructure provided by the Hyak supercomputer system at the University of Washington. Some of the results in this paper were derived using the HEALPix \citep{Gorski2005} package. EWS is currently supported by NASA Postdoctoral Program Fellowship administered by the Universities Space Research Association and the NASA Astrobiology Institute Alternative Earths team under Cooperative Agreement Number NNA15BB03A. TDR is supported by a NASA Exoplanets Research Program award (\#80NSSC18K0349).

\bibliography{ms}
\bibliographystyle{aasjournal}

\end{document}